\documentclass[ %twocolumn,           % Format : preprint, twocolumn
               preprint,
               showpacs,            % Pacs : showpacs, noshowpacs
               letterpaper,             % Size : a4paper, ...
               superscriptaddress,      % Affiliation (Title) : groupedaddress,
               nofootinbib,         % Footnote: footinbib, nofootinbib
               tightenlines,        % Remove additional spaces in a line
               ]{aastex61}

\usepackage{graphicx}% Include figure files
\usepackage{multirow}
\usepackage{bm}% bold math
\usepackage{amsmath}
\usepackage{mathtools}
\usepackage{url}
\usepackage{xcolor}
\DeclareMathOperator\arctanh{arctanh}
\begin{document}

\preprint{APS/123-QED}

\title{Strong gravitational lensing by wave dark matter halos}% Force line
                                % breaks with \\

\author{Antonio Herrera-Mart\'in}%
 \email{a.herrera-martin.1@research.gla.ac.uk}
 \email{antonio.herreramartin@canterbury.ac.nz}
 \affiliation{%
SUPA, University of Glasgow, Glasgow, G12 8QQ, United Kingdom
}%
\affiliation{School of Physical and Chemical Sciences, University of Canterbury
Christchurch, New Zealand.}%Lines break automatically or can be forced with \\
\author{Martin Hendry}
\email{martin.hendry@glasgow.ac.uk}

% \altaffiliation[Also at ]{Physics Department, University of Glasgow.}%Lines break automatically or can be forced with \\
\affiliation{%
SUPA, University of Glasgow, Glasgow, G12 8QQ, United Kingdom
}%

\author{Alma X. Gonzalez-Morales}%
 \email{alma.gonzalez@fisica.ugto.mx}
 \affiliation{Consejo Nacional de Ciencia y Tecnolog\'ia,
Av. Insurgentes Sur 1582. Colonia Cr\'edito Constructor, Del. Benito   Juárez C.P. 03940, M\'exico D.F. M\'exico}
\affiliation{%
Departamento de F\'isica, DCI, Campus Le\'on, Universidad de
Guanajuato, 37150, Le\'on, Guanajuato, M\'exico.}

\author{L. Arturo Ure\~na-L\'opez}
 \email{lurena@ugto.mx}
\affiliation{%
Departamento de F\'isica, DCI, Campus Le\'on, Universidad de
Guanajuato, 37150, Le\'on, Guanajuato, M\'exico.}

\date{\today}% It is always \today, today,

\begin{abstract}
Wave Dark Matter (WaveDM) has recently gained attention as a viable
candidate to account for the dark matter content of the  Universe. In
this paper we explore the extent to which, and under what conditions, dark matter halos in this model are able to reproduce strong lensing systems. First, we explore analytically the lensing properties of the
model, finding that a pure WaveDM density profile, soliton profile,
produces a weaker lensing effect than similar cored profiles. Then
we analyze models with a soliton embedded within an NFW profile, as has
been found in numerical simulations of structure formation. We use a
benchmark model with a boson mass of $m_a=10^{-22} \, {\rm eV}$, for
which we see that there is a bi-modality in the contribution of the
external NFW part of the profile, and some of the free
parameters associated with it are not well constrained. We find
that for configurations with boson masses $10^{-23}$ -- $10^{-22} \, {\rm eV}$, a range of masses preferred by dwarf galaxy kinematics, the soliton
profile alone can fit the data but its size is incompatible with
the luminous extent of the lens galaxies. Likewise, boson masses of
the order of  $10^{-21} \, {\rm eV}$, which would be consistent with
Lyman-$\alpha$ constraints and consist of more compact soliton
configurations, necessarily require the NFW part in order to
reproduce the observed Einstein radii. We then conclude that lens
systems impose a conservative lower bound $m_a > 10^{-24} \, {\rm eV}$ and that the NFW envelope around the soliton must be present to satisfy the observational requirements.
\end{abstract}
             
\section{\label{sec:int}Introduction }

The $\Lambda$CDM model is the most successful theoretical framework in
modern cosmology to explain the process of structure formation in the
Universe on large scales. This model requires the existence of a cold dark
matter (CDM) component that comprises $26\%$ of the total energy
budget, which is best described by a non-relativistic (cold) and
non-interacting fluid~\cite[see][]{Ade:2015xua}.

One of the main predictions from only CDM simulations of structure formation is the appearance of universal cuspy density profiles for the galaxy halos, with the Navarro, Frenk and White (NFW) profile the one most used to describe CDM~\cite[see][]{ONFW}. Despite the successes of CDM at large scales, there are some open questions regarding the  observations on galactic scales, such as: the
``missing satellite problem", the ``cusp core problem", and the
``too-big-to-fail
problem"~\citep[e.g.][]{BurkertDM,Maccio:2012qf,Blok,walterDM,SawalaDM,Klypin:1999uc,toobig},
see also~\cite{Bullock:2017xww} for a recent review. These refer to both theoretical and observational questions on how CDM and barions interplay leads to the shapes, inner density profiles, and abundance of the DM structure at sub-galactic scales.  %Solutions to these problems may come from taking
%into account the effects of baryons in the formation of galaxies, but
%it is doubtful that this is the final answer. 
This also opens the question of weather such observables, or others, can be used to learn more about the dark matter (DM) nature. This has lead to explore the viability of other DM candidates. Actually there is a wide range of DM proposals 
%Another possibility to solve the above mentioned issues is to change the
%paradigm of the nature of dark matter (DM) itself, as has been proposed and explored widely for different candidates 
such as Self-Interacting Dark Matter \citep[see][]{Kaplinghat:2015aga}, Warm Dark Matter \citep[see][]{Gonzalez-Samaniego:2015sfp,Adhikari:2016bei}, Axion/Scalar or Wave Dark Matter ~\cite[e.g.][]{matos:1999et,fuzzy,goodman,Matos:2000ss,bohmer,QMsimpaper,robles2013}, and other specifications of the nature of dark matter particles, which can actually be described in a more general effective theory~\cite[e.g.][]{Cyr-Racine:2015ihg}.

In this paper, our approach is to describe the dark matter as an axion/scalar field that we will refer to as a Wave Dark Matter model (WaveDM, also referred sometimes to as scalar field DM or SFDM, ultralight axion-like DM, fuzzy DM, etc.). This type of  model has been investigated by several other authors ~\cite[e.g.][]{matos:1999et,fuzzy,goodman,Matos:2000ss,bohmer,Lee2017, robles2014,lee2},
and has been found to be able to reproduce the success of the $\Lambda$CDM
model on cosmological scales, but it predicts a natural cut-off on the
mass power spectrum of linear perturbations that could help to alleviate
some of the low-scale issues of
CDM~\cite[e.g.][]{Urena-Lopez:2015gur,Hlozek:2014lca,Matos:2000ss,fuzzy}. Interestingly
enough, all cosmological effects are directly related to a single
parameter, which is the boson mass, $m_a$, of the scalar field particle, although extra observational effects may arise from quartic self-interactions~\cite[e.g.][]{Schive:2017biq,Cedeno:2017sou,Zhang:2017dpp,Zhang:2017flu}. Based
on considering the cut-off of the mass power spectrum, the halo mass
function, the reionization time or the Lyman-$\alpha$ forest, the most
up-to-date constraints suggest that the boson mass must satisfy  $m_a
> 1 \times 10^{-21} \, {\rm eV}$~
\cite[see][]{Irsic:2017yje,Armengaud:2017nkf}.

However, the non-linear process of structure formation under the SFDM hypothesis does not depend on only a single parameter, but instead one requires to take into account at least a second parameter. This fact is indeed considered in many recent studies that try to put constraints on the WaveDM parameters with data coming from, for instance, satellite galaxies in the Milky
Way~\cite[e.g.][]{Bernal:2017oih,Gonzales-Morales:2016mkl,Chen:2016unw,QMpaper,QMsimpaper}. The
aforementioned studies consider that galaxies are described by a
solitonic core with a negligible self-interaction. The soliton solution is just the ground state of the
so-called Schrodinger-Poisson system of
equations~\citep{Ruffini:1969qy,Guzman:2004wj}, and its wave-like
properties provide stability against gravitational collapse -- opening the possibility of naturally-supported, cored halos. The full
prescription of the WaveDM profile requires specification of the boson 
mass $m_a$ together with one of its structure parameters, which can
be taken to be either the central density or the scale radius, while the other is determined by the relation, 
\begin{equation}
    \frac{\rho_s}{M_\odot {\rm pc}^{-3}} = 2.4 \times 10^{12} \left(
    \frac{r_s}{\rm pc} \right)^{-4} \left( \frac{m_a}{10^{-22} \rm eV}
  \right)^{-2} \ . \label{eq:2}
\end{equation}
The boson mass $m_a$ is expected to be a fundamental parameter 
with a single value for all galaxies, while the other two parameters may take values that differ from galaxy to galaxy.
{Hence, it is necessary to think more carefully if we are to obtain meaningful constraints on the boson mass. More specifically, if we consider the boson mass as an universal parameter, on the same footing as any other cosmological parameter, we should certainly be able to use statistical analysis of galaxy data to constrain which values are permitted, as has been proposed in~\cite{SFDMpaper,Diez-Tejedor:2014naa} and more recently carried out in~\cite{Chen:2016unw,Gonzales-Morales:2016mkl}.  However, in general, we may be unable to assert whether there is one single value of $m_a$ that is suitable to satisfy all the possible constraints. For this purpose, in this paper we have selected gravitational lensing as a possible additional tool to study or assess the viability of the WaveDM profile whose parameters are subjected to the constraint in Eq.~\eqref{eq:2}.

Gravitational lensing (strong and weak) has become a powerful astrophysical tool for the study   of the background cosmology, 
structure and substructures of galactic halos.~\cite[see][]{Schneiderbook, cao2015,koopman2005,Cao2016,SlacsI}. In particular, it is possible to extract important information from the stellar kinematics and geometry of strong lensed systems. Furthermore, particular cosmologies where dark energy is not a cosmological constant can be tested by fitting the observed critical lines, Einstein angle and stellar dynamics, given a suitable lens model for the observed systems~\cite[see e.g.][]{futamase2001,Grillo2008,sereno2002,SlacsXI}.
As the geometry of the lensing system can be obtained from image astrometry, it also is a helpful probe for the Hubble parameter and the dark energy contents~\citep{Mitchell2005,biesiada2010}, whereas at the same time it gives information about the structure and formation of Early type galaxies~\citep{SlacsIII,SlacsIV}.}

There are different procedures to extract particular parameters, like the Einstein angle, from the lensing geometry in combination with velocity dispersions from stellar dynamics~\citep{SlacsVII,SlacsX}, where the latter can be obtained along with redshifts from spectroscopic imaging~\citep{ofek2003,ofek2006,Kochanek91,SlacsI,Cao:2011}. When assuming a particular lens model it is usual, as a first approach, to describe strong lenses using an axially symmetric model, the most popular being the Singular Isothermal Sphere (SIS)~\citep{Silviabook,Schneiderbook,SlacsIX,cao2012}. Sometimes, however, there are deviations from the SIS model and the Singular Isothermal Ellipsoid (SIE) is used instead as a non-axially symmetric extension~\citep{SlacsIV,cao2015}. By assuming these models, which consider the total mass distribution of the lens, there is some freedom to test the cosmological parameters~\citep{Cao2016,Jie2016,sereno2002}.
On the other hand, using a fixed cosmology allows one to obtain information of the physical parameters regarding the structure of individual galaxies or clusters of galaxies~\citep{futamase2001,Grillo2008,sereno2002}.
{Furthermore, using information obtained from the strong lensed galaxies is possible to test the gravitational weak field, through the post-Newtonian parameter $\gamma$, where the parameterization of the profiles to describe  are sensitive to the total mass distribution. This can be used to test the validity of General Relativity on large scales. Recently, it has been found that for at least $\sim 2$kpc, the lensing galaxies profiles agree with General Relativity~\cite[see e.g.][]{cao2018,Collett1342}}.

For the description of lensing systems, several mass distribution profiles exist which have been successful enough to represent observed data; the most popular, which include the SIS and SIE~\cite[see e.g.][]{keeton2001}, follow a power-law distribution. These  are very successful to describe the observed data for lensing geometry and stellar dynamics when considering the total matter contents (luminous+dark matter)~\citep{suyu2014,SlacsV,SlacsVII,SlacsIX,cao2012,cao2015}.  {It is important to stress out that these modelling choices does not impact on the estimated Einstein angle of the lens.} Nevertheless, these modelling choices do not explicitly give much information about the properties of the internal structure of the dark matter distribution in the galaxies. 
For this reason, other composite models where the baryonic matter and the dark matter are treated separately are also used~\cite[e.g.][]{suyu2014,ONFW,Burkerpaper}. The luminous part,
which contains the baryonic matter, is described by a SIS, but it is common to use Sersic or more general power-law luminosity profiles~\citep{cardone2004,Cao2016}. The most popular to describe the DM halo of a galaxy is the NFW density profile~\citep{NFWpaper,Bartelmann}, which introduces a family of generalized NFW profiles~\citep{Wyithe2001}. Other popular profiles are the Burkert profile~\citep{Burkerpaper,BurkertDM}, and cuspy halo models~\citep{keeton2001,munoz2001}. It has been shown that the NFW profile correctly describes the observed lensing signal in large samples of galaxies~\citep{SlacsIV} and clusters of galaxies~\citep{Niikura2015}.
 
Since the WaveDM model is considered a feasible candidate for DM, in this work we study the behavior of, and put constraints upon, a WaveDM type of profile acting as a single galactic gravitational lens, and we obtain the conditions under which the profile will be able to produce strong lensing. As we shall see, the WaveDM profile consists of a solitonic core plus a tail in the outer parts that follows the prescription of an NFW profile, whose properties are closely interlinked by their matching conditions such that a soliton is always present in the centre of the DM halo. In this respect, we are not interested in the individual properties of the single profiles (soliton vs NFW), but rather on the conditions under which the complete WaveDM profile could be consistent with lensing data.
 
The remainder of this paper is organized as follows. The basic lensing equations for any given density profile are described in
Sec.~\ref{lensingQM}, where we also introduce the explicit lensing expressions for the particular case of the WaveDM profile.  In Sec.~\ref{sec:dat} we describe our statistical analysis and present the results arising from the comparison of the WaveDM model predictions with 
selected data from the SLACS catalog. Finally, the general conclusions are presented in Sec.~\ref{sec:dis}. Some analytic solutions of the lens equations used in the text are shown in the appendix.

%-----------------------------------------
\section{Gravitational lensing with a $\psi$DM profile}
\label{lensingQM}
%-----------------------------------------

\subsection{General lensing equations \label{sec:lensing-equations-}}
One of the main predictions from Einstein's General Relativity is
the bending of light as it passes close to a massive body. The deflection angle produced by this effect depends on the mass of the deflector, which then acts like a lens.  This deflector may be approximated by a point-like mass, like in the case of a star, but for more massive objects like galaxies it is better to represent them as extended masses which are described by their density profiles. 
For the purposes of this work, we shall consider  galactic lenses, and therefore the density profile described in this section will be representing one galaxy acting as a single lens.

The simplest type of lens is a system with a point mass $M$ located close to the line of sight to a luminous source $S$. Due to the gravitational field of the point mass, a light ray is deflected in its path to the observer; this is described by the lens equation in the thin-lens approximation. The same approximation also holds for a mass distribution, 
in which case the lens equation is~\citep{2002glml.book.....M},  
\begin{equation}
  \beta = \theta - \frac{m(\theta)}{\pi \Sigma_{cr} D^2_{OL}\,
    \theta} \, , \label{eq:10c}
\end{equation}
that relates the (unobservable) angle between the line of sight and the path from the observer to the actual position of the source, $\beta$, and to the apparent position of the source (the image), $\theta$, and the mass distribution that is causing the lensing  $m(\theta)$.Also, $D_{OL}$ is the angular distance from the observer to the lens, that we denote by the subindex (OL). Here we have  assumed that $m(\theta)$ is the projected mass enclosed in a circle of radius $\xi \equiv D_{OL} \theta$; 
more explicitly, we can write

\begin{subequations} 
\begin{equation}
  \label{eq:10a}
  m(\xi) = 2\pi \int^\xi_0 d\hat{\xi} \, \hat{\xi} \, \Sigma(\hat{\xi})
  \, .
\end{equation}
The projected surface mass density $\Sigma(\xi)$ can be calculated 
directly from the (spherically symmetric) density profile $\rho(r)$ of the 
lensing object as:
\begin{equation}
  \label{eq:10b}
  \Sigma(\xi) = 2 \int^{z_{\rm max}}_0 dz \, \rho(z,\xi) \, ,
\end{equation}
\end{subequations}
where $z \equiv \sqrt{r^2 - \xi^2}$ is a coordinate orthogonal to the line of sight, so that $0 \leq \xi \leq r$. If the
lens system has a finite radius $r_{\rm max}$, then $z_{\rm max} =
\sqrt{r^2_{\rm max} - \xi}$; otherwise, we can put $z_{\rm max} \to
\infty$ in the integral given by Eq.~\eqref{eq:10b}.

Let us consider the case in which the density profile $\rho(r)$ has
a characteristic density $\rho_s$, and a characteristic radius
$r_s$, such that $\rho(r) = \rho_s f(r/r_s)$, where $f$ is the function that accounts for the shape of the profile. We can then write Eq.~\eqref{eq:10c} in the dimensionless form
\begin{equation}
 \beta_*(\theta_*) = \theta_* - \lambda \frac{m_\ast(\theta_*)}{\theta_*}
 \, , \label{eq:19a}
\end{equation}
where the different distances are normalized in terms of $r_s$:
$\beta_* = D_{OL} \beta/r_s$, $\theta_* =  D_{OL} \theta/r_s$, and
then $\xi_\ast = \xi/r_s = \theta_\ast$. The latter equation means that 
the normalized variables $\xi_\ast$ and $\theta_\ast$ can be used
interchangeably, and then hereafter we will use $\theta_\ast$ as our
distance variable\footnote{For the sake of simplicity in the notation, we are using the same same angular variables (together with an asterisk) to denote the new normalized distances.}. Likewise, the total mass, as given in 
Eq.~\eqref{eq:10a}, is normalized as
\begin{subequations}
\begin{equation}
  m_\ast(\theta_*) = \frac{m(\theta_*)}{\rho_{\rm s} r^3_{\rm s}} =
  2\pi \int^{\theta_\ast}_0 d\hat{\theta}_\ast \, \hat{\theta}_\ast \,
  \Sigma_\ast (\hat{\theta}_\ast) \, . \label{eq:14}
\end{equation}
The normalized projected surface mass density, from Eq.~\eqref{eq:10b}, is
\begin{equation}
  \Sigma_\ast (\theta_\ast) = \frac{\Sigma(\theta_*)}{\rho_{\rm s}
    r_{\rm s}} = 2 \int^{z_{\rm max \ast}}_0 dz \, f(z,\theta_\ast) \,
  , \label{eq:18}
\end{equation}
\end{subequations}
with $z= \sqrt{r^2_\ast - \theta^2_\ast}$ and $r_\ast = r/r_s$. The new
parameter $\lambda$ in Eq.~\eqref{eq:19a} is then given by
\begin{equation}
\label{eq:lenseq}
 \lambda \equiv \frac{\rho_{\rm s}r_{\rm s}}{\pi \Sigma_{\rm cr}} = 10^{-3}
 \frac{0.57}{h} \left( \frac{\rho_{\rm s} r_s}{M_\odot {\rm pc}^{-2}}
 \right) \frac{d_{\rm OL} d_{\rm LS}}{d_{\rm OS}} \, ,
\end{equation}
where we have defined the reduced (dimensionless) angular distances $d_A = D_A H_0/c$. The angular diameter distance $D_A$ as a function of redshift is computed in the standard way~\cite[see e.g.][]{Hogg1999}, assuming cosmological model parameters as given by the Planck 2015 results~\citep{Ade:2015xua}. Equation~\eqref{eq:lenseq} contains information about the lensing properties of any given model, together with that of the different distances involved in the lens system, namely  that between the observer and the lens (OL), the observer and the source (OS), and the lens and the source (LS).\footnote{This is the same parameter used in~\cite{SFDMpaper}, but also see~\cite{Burkerpaper} in which the definition of $\lambda$ differs by a factor of $1/4\pi$.}

One particular case of interest is that of perfect alignment between
the luminous source and the lens system for which
$\beta_*(\theta_{*E}) = 0$. This in turn defines an Einstein ring with
radius $R_E = D_{OL} \theta_E$, and an associated angular radius
$\theta_E$. In terms of our normalized variables, we see that the normalized angular Einstein radius $\theta_{\ast E}$ directly is the ratio of the Einstein radius to the (characteristic) scale radius of the density profile, $\theta_{\ast E} = R_E/r_s $. Moreover, the angular radius $\theta_{*E}$ must also be a solution of the equation [see Eq.~\eqref{eq:19a}]
\begin{equation}
 \lambda = \frac{\theta^2_{\ast E}}{m_\ast(\theta_{\ast E})} \,
 . \label{eq:11}
\end{equation}

\begin{figure}[tp!]
  \centering
    \includegraphics[width=0.49\textwidth]{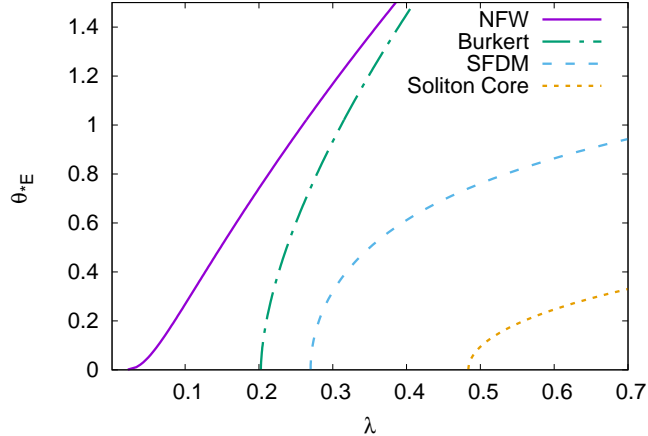}
     \caption{Normalized Einstein radius $\theta_{\ast E}$ as a function of 
	      $\lambda$ for different density profiles, see
              Eq.~\eqref{eq:11}. The point where each curve crosses the
              horizontal axis indicates the (intrinsic) critical value
              $\lambda_{\rm cr}$ for each profile, in agreement with
              the values calculated from Eq.~\eqref{eq:15b} as shown in Table~\ref{tab:1}.} 
    \label{fig:Lambda}
\end{figure}

Interestingly enough, Eq.~\eqref{eq:11} shows that the lensing properties of a system with a density profile of the form
$\rho(r) = \rho_s f(r/r_s)$ are independent of the density and
distance scales, and are mostly sensitive to the particular shape of
the density profile. The physical parameters of the system are then
concentrated in the dimensionless parameter $\lambda$ in
Eq.~\eqref{eq:lenseq}, and the latter can be calculated from
Eq.~\eqref{eq:11} without any prior knowledge of the given physical scales in the system, namely $\rho_s$ and $r_s$, under only the assumption of perfect alignment (see Fig.~\ref{fig:Lambda} below for
an example).
\begin{table}[tp!]
 \begin{center}
 \renewcommand{\arraystretch}{2}
   \begin{tabular}{| l |c@{\hspace{10pt}} | c | c |}
 \hline
  Name  & Density profile $f(r)$ & $\lambda_{cr}$ & References \\ \hline
  NFW & $\left[ (r/r_s) (1 + r/r_s)^2 \right]^{-1}$
  & 0 & \cite{NFWpaper} \\ 
  Burkert  & $\left[ (1 + r/r_s) (1 + ^2r/r^2_s) \right]$ & $2/\pi^2 \simeq 0.203$ & \cite{Burkerpaper}  \\  
  SFDM  & $\sin(\pi r/r_s)/(\pi r/r_s)$ &
  $0.27 $ & \cite{SFDMpaper} \\ 
  WaveDM & $( 1 + r^2/r^2_s)^{-8}$ &
  $\frac{2048}{429 \pi^2} \simeq 0.484$ & \citep{Marsh:2015wka,QMpaper} \\ \hline  
 \end{tabular}
 \caption{The intrinsic value $\lambda_{cr}$, calculated from
   Eq.~\eqref{eq:15b} for dark matter halos with different density
   profiles.} %This critical value is the minimum value of $\lambda$ in
  % Eq.~\eqref{eq:11} for which the lens system will produce multiple 
  % images, i.e. an Einstein ring.}   
 \label{tab:1}
\end{center}
%\tablerefs{(1)~, (2)~, (3)~}
\end{table}

There is a critical value $\lambda_{\rm cr}$ that is the
smallest value of $\lambda$ for which an Einstein ring appears, which must correspond to the limit $\theta_{*E} \to 0$ in
Eq.~\eqref{eq:11}. As we shall show now, such a critical value can be calculated analytically in the general case. To avoid the divergence
at $\theta_{\ast E} =0$ (where $m_\ast(0)=0$), we make use of the L'H\^{o}pital
rule in Eq.~\eqref{eq:11}, and from Eq.~\eqref{eq:10a} we finally
obtain
\begin{equation}
  \label{eq:15b}
  \lambda^{-1}_{\rm cr} = \pi \Sigma_\ast(0) = 2 \pi \int^{r_{\rm max
      \ast}}_0 d\hat{r}_\ast \, f(\hat{r}_\ast) \, ,
\end{equation}
where $\Sigma(0)$ is the central value of the projected surface
mass density given by Eq.~\eqref{eq:10b}. 
Eq.~\eqref{eq:15b} is quite a simple formula
for the calculation of $\lambda_{\rm cr}$ for any given density
profile $\rho(r)$.\footnote{It should be noted that the definition of
  $\lambda_{\rm cr}$ depends on the chosen scale radius for normalization $r_s$, so that the value obtained from Eq.~\eqref{eq:15b} in our case is considering that $r_s$ coincides with the intrinsic distance scale in the density profile $\rho(r)$.}

As said before, Eq.~\eqref{eq:15b} suggests that the critical
value $\lambda_{\rm cr}$ just depends on the particular shape of the
given density profile and no information is necessary about its
other physical parameters. The values of $\lambda_{\rm crit}$, calculated
from Eq.~\eqref{eq:15b} for density profiles that are well-known in the
literature, are shown in Table~\ref{tab:1}. For these profiles we also
show in Fig.~\ref{fig:Lambda} the Einstein angle $\theta_{\ast
  E}$ as calculated from Eq.~\eqref{eq:11}. As expected, the Einstein
angle is the smallest for the WaveDM profile (Eq.~\eqref{eq:solprofile})
alone, which also means that it is the one with the weakest lensing
signal.

We should mention here an additional use of Eq.~\eqref{eq:11} to constrain the free parameters of a given
density profile. It relates to the fact that any DM halo
characterized by a particular density profile needs to satisfy the
constraint $\lambda \geq \lambda_{\rm cr}$ if it is to produce a
lensing signal. Using Eqs.~\eqref{eq:lenseq} and~\eqref{eq:15b}, the
latter statement can be re-written as
\begin{equation}
  \frac{\rho_s r_s}{M_\odot {\rm pc}^{-2}} \geq 10^3 \frac{h}{0.57}
  \frac{d_{\rm OS}}{d_{\rm OL} d_{\rm LS}} \lambda_{\rm cr} \,
  . \label{eq:10}
\end{equation}
Equation~\eqref{eq:10} establishes a minimum value for the (structural)
surface density $\rho_s r_s$ of any given DM profile in terms of the
measured quantities of a lens system. Although the
constraint Eq.~\eqref{eq:10} is satisfied automatically by the NFW
profile, for which $\lambda_{\rm crit} =0$, this is not the case for
the other profiles listed in Table~\ref{tab:1}.

\subsection{\label{sec:den} Combined density profile of WaveDM}
For the density profile of WaveDM halos we will
consider the model described in~\cite{QMsimpaper,QMpaper},
which arises from the study of extensive N-body simulations. The
profile consists basically of two parts: one part describing a core
sustained by the quantum pressure of the boson particles, also known
as the soliton profile, and another part that resembles a NFW-like profile
in the outer parts of the halo. As argued in~\cite{marsh}, the
transition at some radius to a NFW profile must be expected from the
change of behavior to CDM on scales larger than the natural length of
coherence, which should be proportional to the associated Compton
length of the boson particles (in full units, the Compton length is $L_C = \hbar / (mc)$, where $\hbar$ is the (reduced) Planck's constant and $c$ is the speed of light). 

The soliton profile is given by
\begin{equation}
 \rho_{\rm sol} (r) = \frac{\rho_s}{(1+ r^2/r^2_s)^8} \,
 , \label{eq:solprofile}
\end{equation}
where $r_s$ and $\rho_s$ are its characteristic
radius and central density contrast, respectively. This profile was first studied in detail in~\cite{QMpaper}, although here we are following the nomenclature adopted in~\cite{marsh}, where it is also shown 
that the profile fits well the ground-state solution of the so-called Schr\"{o}dinger-Poisson (SP) system
of equations~\citep{Ruffini:1969qy,Guzman:2004wj}. In this
respect, the soliton profile is strongly related to the wave
properties (via the Schr\"{o}dinger equation) of the boson particles.

One important property of the profile given in Eq.~\eqref{eq:solprofile} 
is that it must also obey the intrinsic scaling symmetry of the SP
system~\citep{Guzman:2004wj}. If $0 < \hat{\lambda} \ll 1$ is a constant
parameter, it can be shown that the central density and radius in the
soliton profile are given by
\begin{equation}
  \rho_s = \hat{\lambda}^4 m^2_a \, m^2_{\rm Pl}/4\pi \, ,
  \quad r_s = (0.23 \, \hat{\lambda} \, m_a)^{-1} \, , \label{eq:1}
\end{equation}
This equation suggests that the intrinsic, physical, quantities of the
soliton profile in Eq.~\eqref{eq:solprofile} are related as shown in Eq.~\eqref{eq:2}. This relation will be important later when we discuss
the constraints on the boson mass $m_a$.
%, which we rewrite here just for clarity:
%\begin{equation}
%    \frac{\rho_s}{M_\odot {\rm pc}^{-3}} = 2.4 \times 10^{12} \left(
%    \frac{r_s}{\rm pc} \right)^{-4} \left( \frac{m_a}{10^{-22} \rm eV}
%  \right)^{-2} \, \nonumber.
%  % \label{eq:2}
%\end{equation}

For the NFW profile at the outskirts of the galaxy halo we adopt the
following parametrization
\begin{equation}
 \rho_{\rm NFW}(r) = \frac{\rho_s \, \rho_{\rm NFW
     \ast}}{\alpha_{\rm NFW} \, (r/r_s) \left(1+ \alpha_{\rm NFW} \,
   r/r_s \right)^2} \, . \label{eq:NFWprofile}
\end{equation}
Notice that in writing Eq.~\eqref{eq:NFWprofile} we are assuming the
following implicit definitions for the scale radius and density,
respectively, of the NFW profile: $r_{\rm NFW} = r_s/\alpha_{\rm NFW}$
and $\rho_{\rm NFW} = \rho_s \, \rho_{\rm NFW \ast}$, where
both $\alpha_{\rm NFW}$ and $\rho_{\rm NFW \ast}$ are dimensionless
numbers.

Unfortunately, there is not precise information in~\cite{QMsimpaper}
about the transition in a galaxy halo from the soliton
profile of Eq.~\eqref{eq:solprofile} to the NFW profile of Eq.~\eqref{eq:NFWprofile} 
in the general case.  Hence, for the present work we adopt the
convention for a combined profile as suggested in~\cite{marsh}
\begin{equation}
  \rho(r) = \Theta(r_\epsilon-r)\rho_{\rm sol}(r)+\Theta(r-r_\epsilon) \rho_{\rm NFW}(r) \, .
  \label{eq:profile}
\end{equation}
where $\Theta(r_\epsilon-r)$ is the Heaviside step function. 
Here, $r_\epsilon$
is the matching radius where the transition between the individual
profiles occurs, and which satisfies the condition
$\rho(r_\epsilon) = \epsilon \rho_s$. Notice that $0 <
\epsilon < 1$ if the transition between the profiles is to occur at
the outskirts of the galaxy halo.

In general terms, and under our parametrization, there are six free
parameters in the combined profile (Eq.~\eqref{eq:profile}):
$(\rho_s,r_s, \rho_{\rm NFW \ast},
\epsilon,r_\epsilon,\alpha_{\rm NFW})$. We will now derive two new
constraints that arise from the continuity of the combined density
profile at the matching radius which will help us to reduce the number
of free parameters.

For a continuous density function, we must impose the condition
\begin{equation}
\rho_{\rm sol}(r_\epsilon) = \epsilon \rho_s = \rho_{\rm
  NFW}(r_\epsilon) \, . \label{eq:3}
\end{equation}
When Eq.~\eqref{eq:3} is applied to the soliton
profile of Eq.~\eqref{eq:solprofile}, we obtain
\begin{subequations}
  \label{eq:4}
\begin{equation}
  \label{eq:4a}
  r_{\epsilon \ast} = r_\epsilon/r_s = \left( \epsilon^{-1/8} - 1 \right)^{1/2} \, ,
\end{equation}
which basically establishes the interchangeability of the (dimensionless) matching radius $r_{\epsilon \ast}$ and $\epsilon$. In the case of the NFW
profile (Eq.~\eqref{eq:NFWprofile}), the continuity condition (Eq.~\eqref{eq:3}) establishes that
\begin{equation}
  \label{eq:4b}
  \epsilon^{-1} \rho_{\rm NFW \ast} = \alpha_{\rm NFW} \, r_{\epsilon
    \ast} \left( 1 + \alpha_{\rm NFW} r_{\epsilon \ast} \right)^2 \, ,
\end{equation}
which, taking into account Eq.~\eqref{eq:4a}, can be written as
\begin{equation}
  \label{eq:6}
  \rho_{\rm NFW \ast} = \frac{ \alpha_{\rm NFW} \, r_{\epsilon \ast}
    \left( 1 + \alpha_{\rm NFW} \, r_{\epsilon \ast} \right)^2}{\left(
      1+ r^2_{\epsilon \ast} \right)^8} \, .
\end{equation}
\end{subequations}
Equation~\eqref{eq:6} indicates the (normalized) density $\rho_{\rm
  NFW \ast}$ that is required for a correct matching between the
soliton and NFW profiles, for given values of $\alpha_{\rm NFW}$ and
$r_{\epsilon \ast}$.

\begin{figure*}[thp!]
\centering
\includegraphics[width=0.49\textwidth]{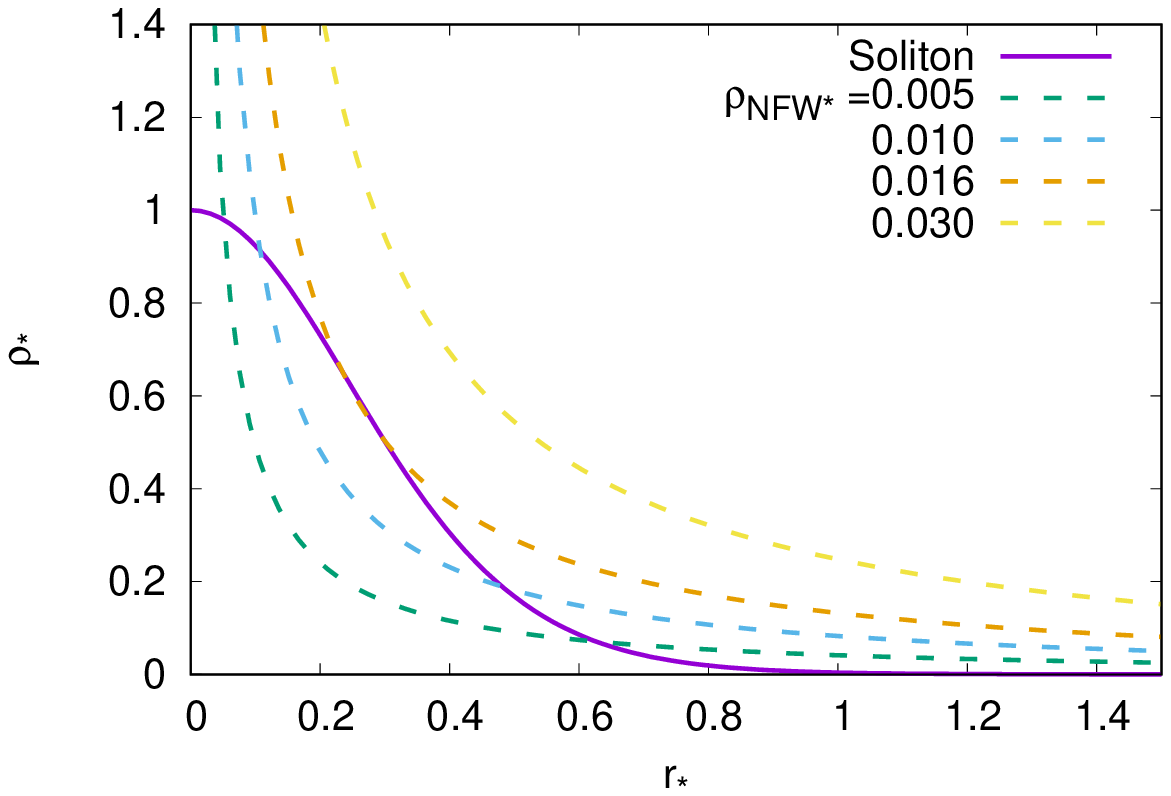}
\includegraphics[width=0.49\textwidth]{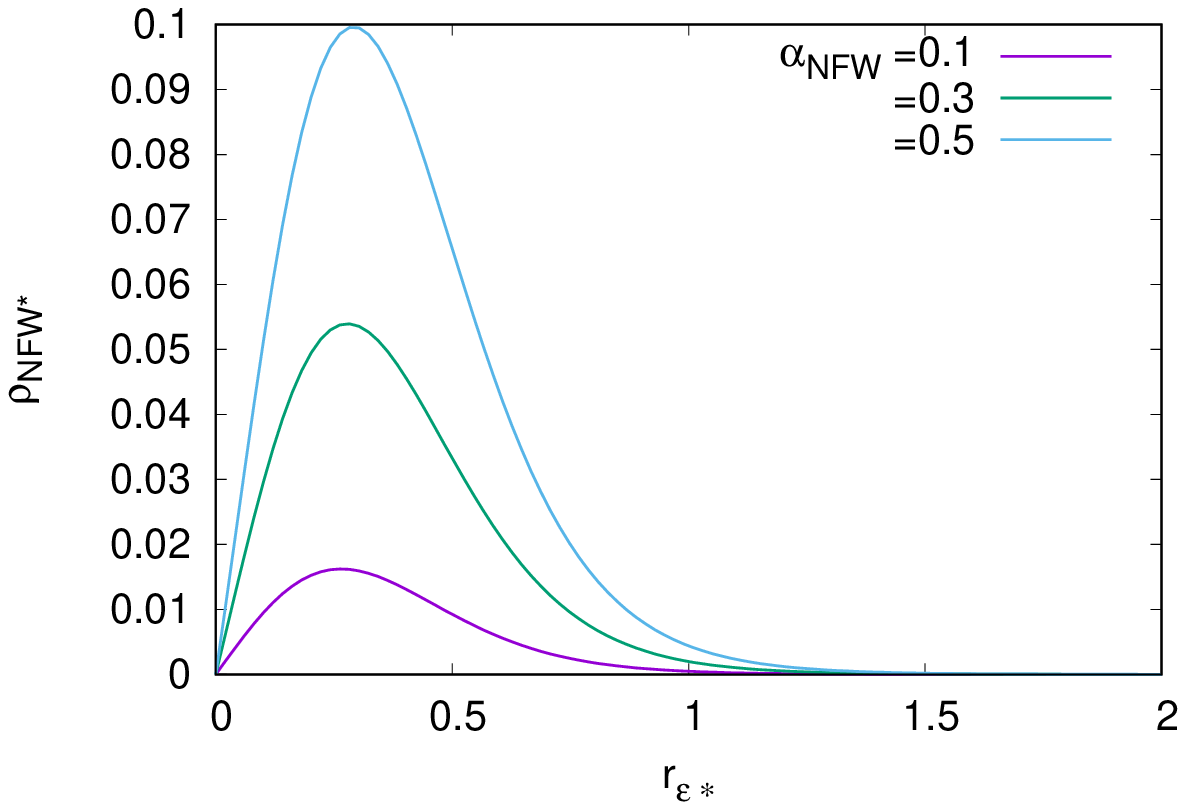}
\caption{\label{fig:0a} Determination of the matching radius, $r_{\epsilon \ast}$, for the density profile in Eq.~\eqref{eq:3}.  (Left) Normalized soliton and  NFW 
 density profiles showing there are at most two possible for $r_{\epsilon \ast}$ for each combination of $\alpha_{\rm NFW}$ (set to 0.1 only for illustrative purposes) and $\rho_{\rm NFW \ast}$. (Right)
NFW normalizing density factor $\rho_{\rm NFW \ast}$ as a function of $r_{\epsilon \ast}$ for different values of $\alpha_{\rm NFW}$, as obtained from Eq.~\eqref{eq:6}.  }
\end{figure*}

However, one can see that the continuity constraint (Eq.~\eqref{eq:6})
actually shows a hidden degeneracy: once the values of $\alpha_{\rm
  NFW}$ and $\rho_{\rm NFW \ast}$ are fixed, there can be up to two
solutions for the matching radius $r_{\epsilon \ast}$. This is a direct
consequence of the fact that the crossing of the density profiles (Eqs.~\eqref{eq:solprofile} and~\eqref{eq:NFWprofile}) can occur at
most at two different points, as illustrated in the left-hand panel of Fig.~\ref{fig:0a}, which shows normalized density profiles for 
$\alpha_{NFW} = 0.1$ and different values of the normalized density $\rho_{\rm NFW \ast}$.

Fig.~\ref{fig:0a} (left panel) also shows that there exists a maximum value of $\rho_{\rm  NFW \ast}$ beyond which the profiles do not cross each other. This fact can be understood in terms of Eq.~\eqref{eq:6}, which we evaluate for different values of $\alpha_{\rm NFW}$ in the right-hand panel of Fig.~\ref{fig:0a}. Here we can see that, for each 
$\alpha_{\rm NFW}$, there is always a maximum value of
$\rho_{\rm NFW \ast}$, as a function of the matching radius $r_{\epsilon \ast}$, that corresponds to the case in which the soliton
and NFW density profiles barely touch, as seen in the left-hand panel of
Fig.~\ref{fig:0a}.

To avoid the hidden degeneracy, and to select always a combined profile with an interior soliton shape, we will choose those cases for which
$r_{\epsilon \ast} \geq r_{\rm \epsilon \ast, max}$, where $r_{\rm
  \epsilon \ast, max}$ is the matching radius corresponding to the
maximum value of $\rho_{\rm NFW \ast}$. A straightforward calculation
from Eq.~\eqref{eq:6} shows that $r_{\rm \epsilon \ast, max}$ is a
root of the cubic equation
\begin{equation}
  13 \alpha_{\rm NFW} r^3_{\rm \epsilon \ast, max} + 15 r^2_{\rm
    \epsilon \ast, max} - 3 \alpha_{\rm NFW} r_{\rm \epsilon \ast,
    max} = 1 \, . \label{eq:16}
\end{equation}
%\end{widetext}
Although there is a general solution to this equation, it can be shown that the limits
for small and large values of $\alpha_{\rm NFW}$ are
\begin{subequations}
  \label{eq:8}
\begin{eqnarray}
  \lim_{\alpha_{\rm NFW} \to 0} r_{\rm \epsilon \ast, max} &=& (1/\sqrt{15}) \,
                                                               , \label{eq:8aa}
  \\
  \lim_{\alpha_{\rm NFW} \to \infty} r_{\rm \epsilon \ast, max} &=& (\sqrt{3/13}) \, . \label{eq:8bb}
\end{eqnarray}
\end{subequations}
This means that in absolute terms the maximum value of $\rho_{\rm NFW
  \ast}$ must be located in the range $ 0.25 < r_{\rm \epsilon \ast,
  max} < 0.48$, which is in agreement with the values observed in the
right-hand panel of Fig.~\ref{fig:0a}. After all this, it is possible to reduce the number of free parameters that describe the combined profile~\eqref{eq:profile} to only four: $\rho_s$, $r_s$, $r_\epsilon$ and $\alpha_{\rm NFW}$. By means of these parameters and the constraints discussed above, the other parameters are fully specified. 

Notice that our chosen normalization
is such that the physical parameters in the NFW
profile (Eq.~\eqref{eq:NFWprofile}) are given in terms of those in the
soliton profile (Eq.~\eqref{eq:solprofile}). This means, for instance, that
$\rho_{\rm NFW \ast} > 1$ ($\rho_{\rm NFW \ast} < 1$) is equivalent to
$\rho_{\rm NFW} > \rho_s$ ($\rho_{\rm NFW} < \rho_s$), whatever the
physical value of $\rho_s$ is. Likewise, we find that $\alpha_{\rm
  NFW} < 1$ ( $\alpha_{\rm NFW} > 1$) corresponds to $r_{\rm NFW} >
r_s$ ($r_{\rm NFW} < r_s$), even if the physical value of $r_s$ is not
known beforehand. The same will apply for the matching radius, since
$r_{\epsilon \ast} > 1$ ($r_{\epsilon \ast} < 1$) means that matching
occurs beyond the soliton radius and then $r_\epsilon > r_s$
(before the soliton radius, and then $r_\epsilon < r_s$).

It must be noticed also that the prescription above for the matching of the density profiles in Eq.~\eqref{eq:profile} means that the NFW part is always subjected to the presence of the central soliton. For instance, the NFW profile can be diluted away if the matching radius $r_{\epsilon \ast} \to \infty$ (which also means that $\rho_{\rm NFW \ast} \to 0$), and then the density profile becomes the soliton one alone, $\rho(r) \simeq \rho_{\rm sol}(r)$. On the other hand, if $r_s \to 0$ the central soliton becomes small but much more massive and more dense, because of the scaling symmetry shown in Eq.~\eqref{eq:2}, so that it dominates the matter contents over that of the NFW profile. The conclusion here is that under our parametrization the density profile (Eq.~\eqref{eq:profile}) can become the soliton profile only if $r_{\epsilon \ast} \to \infty$, but it is not possible to do the same for the NFW part; in this sense, the complete profile (Eq.~\eqref{eq:profile}) should always be seen as that of a central soliton with a subdominant NFW tail.

We want to stress that the complete profile~\eqref{eq:profile} should not be confused with the so-called cored NFW profile that exists already in the literature. The latter is of the form $\rho(r) \sim (r/r_s)^{-\beta}/(1+r/r_s)^2$ with $0 <\beta < 1$, and whose lensing properties have been analyzed in~\cite{Wyithe2001}. A comparison of the lenses produced by the cored-NFW profile and the WaveDM one is beyond the purpose of this work, as our primary intention is to constrain the parameters in Eq.~\eqref{eq:profile} and to obtain from them credible bounds on the mass of the boson particles.

As a final note, we emphasize the convenience of the chosen parametrization in terms of the soliton characteristic quantities, as the soliton and NFW parameters must follow well defined scaling constraints that are intrinsic to the WaveDM. These scaling properties will then be already explicit in the complete profile~\eqref{eq:profile} when making a comparison of the model with lensing data.

\subsection{\label{sec:grav} Gravitational Lensing}
To obtain the lensing properties of the combined
profile given by Eq.~\eqref{eq:profile}, we follow the recipe described in Sec.~\ref{sec:lensing-equations-}.  We first need to compute
the projected surface mass density (Eq.~\eqref{eq:10b}). Because of the presence of the step functions in Eq.~\eqref{eq:profile}, the
integral in Eq.~\eqref{eq:10b} naturally separates as
%\begin{widetext}
\begin{equation}
\Sigma_\ast(\theta_\ast,r_{\epsilon \ast},\alpha_{\rm NFW}) = 2 
\begin{dcases}
\int \limits^{\sqrt{r^2_{\epsilon \ast}-\theta^2_\ast}}_0 \frac{dz}{\left( 1
    + \hat{r}^2 \right)^8} + \frac{r_{\epsilon \ast} \left( 1 +
    \alpha_{\rm NFW} \, r_{\epsilon \ast} \right)^2}{\left( 1+
    r^2_{\epsilon \ast} \right)^8} \int
\limits^\infty_{\sqrt{r^2_{\epsilon \ast} - \theta^2_\ast}}
\frac{dz}{\hat{r} \left( 1+ \alpha_{\rm NFW} \, \hat{r} \right)^2} \,
, & \theta_\ast < r_{\epsilon \ast} \, , \\
    \frac{r_{\epsilon
    \ast} \left( 1 + \alpha_{\rm NFW} \, r_{\epsilon \ast}
  \right)^2}{\left( 1+ r^2_{\epsilon \ast}
  \right)^8} \int^\infty_0
      \frac{dz}{\hat{r} \left( 1+ \alpha_{\rm NFW} \, \hat{r}
        \right)^2} \, , & \theta_\ast \geq r_{\epsilon \ast} \, .
    \end{dcases} \, . \label{eq:8a}
\end{equation}
 %\end{widetext}
It should be understood that the integrals in Eq.~\eqref{eq:8a} are done
along the line of sight. Notice also that we are following our convention in Sec.~\ref{lensingQM} for normalized quantities, namely $\Sigma_* = \Sigma/(\rho_s r_s)$, $\theta_\ast = \xi/r_s$ and $z = \sqrt{r^2_\ast - \theta^2_\ast}$. The analytical expressions for the integrals in Eq.~\eqref{eq:8a} can be found in appendix~\ref{sec:integral-solutions-}.

Equation~\eqref{eq:8a} shows that the projected
surface mass density only depends upon the characteristic radii and densities of the combined density profile (Eq.~\eqref{eq:profile}). For instance, if we keep $r_s$ fixed, it can be shown that
\begin{equation}
  \lim_{r_{\epsilon \ast} \to \infty} \Sigma_\ast (\theta_*, r_{\epsilon
    \ast}, \alpha_{\rm NFW}) = 0.658 \left( 1 + \theta_\ast^2
    \right)^{-15/2} \, , \label{eq:14a}
\end{equation}
a result that is obtained from the first branch in
Eq.~\eqref{eq:8a}. Notice that Eq.~\eqref{eq:14a} is exactly the
result for the soliton profile (Eq.~\eqref{eq:solprofile}) alone. Also, as we have mentioned before, it is not possible to recover the standard result of the surface density for the NFW profile by letting $r_s \to 0$, as in this case the matter content is still dominated by the central soliton.

%we can recover, from the second branch in Eq.~\eqref{eq:8a}, the  which is equivalent to $\alpha_{\rm NFW} \to 0$ (with $r_{\rm NFW} = {\rm const.}$) and $r_{\epsilon \ast} \to 1/\sqrt[]{15}$, see Eq.~\eqref{eq:8aa}. Thus, under our parametrization of the complete density profile~\eqref{eq:profile} the limit $\alpha_{\rm NFW} \to 0$ will indicate the dominance of the NFW profile over the soliton one.}
%we cannot recover the result of the NFW profile if $r_{\epsilon \ast} \to 0$, as the second branch in Eq.~\eqref{eq:14a} indicates that $\Sigma_\ast \to 0$ in such a case {\bf We have to explain here that this limit is not well defined mathematically because of the $r_{max}$ constraint.}. In addition, it must be remembered that the operation $r_{\epsilon \ast} \to 0$ is not permitted by the constraint $r_{\epsilon \ast} \geq r_{\epsilon \ast, \rm max}$, see Eq.~\eqref{eq:16}.

\begin{figure*}[htp!]
  \centering
    \includegraphics[width=0.49\textwidth]{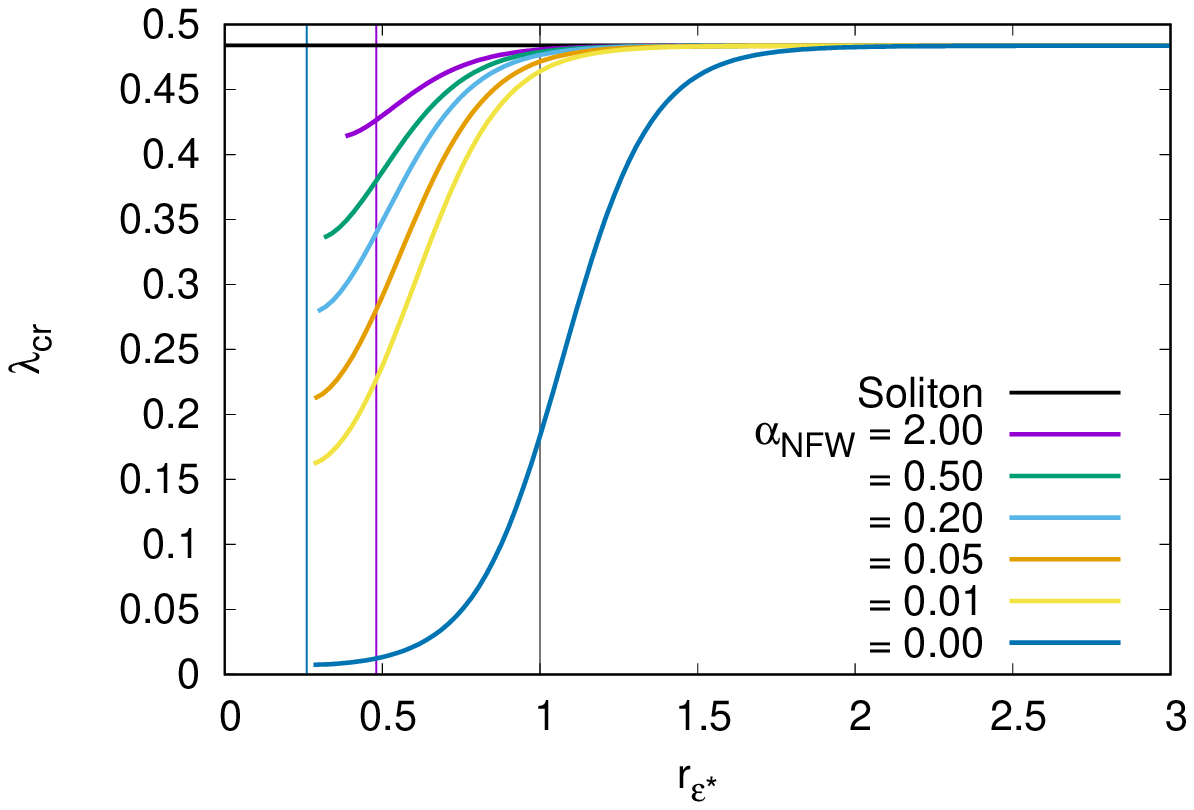}
    \includegraphics[width=0.49\textwidth]{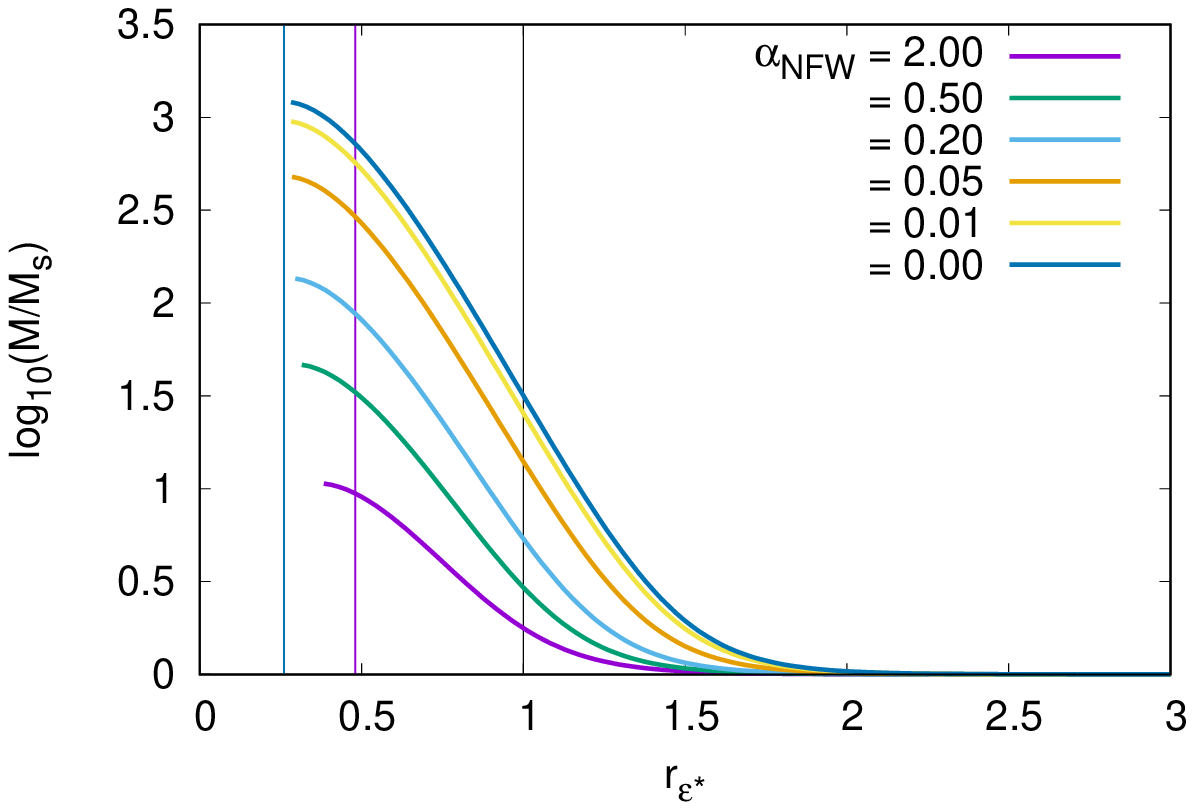}
        \caption{  {Critical value $\lambda_{\rm cr}$ to produce strong lensing (left). Total lens mass $M$ normalized by the soliton mass $M_s$ (right). Both as a function of the matching radius $r_{\epsilon \ast}$ for different density profiles characterized by $\alpha_{\rm NFW}$. The pure soliton case,  $\lambda_{\rm cr} \simeq 0.48$ is recovered asymptotically in the limit $r_{\epsilon \ast} \to \infty$. Vertical lines marks the position of the normalized soliton radius at $r_{\epsilon \ast} = 1,0.5,0.25$ (gray,purple and blue). The presence of the NFW part ease the formation of multiple images. See the text for more details. }
       }
%     \caption{(Left) The critical value $\lambda_{\rm cr}$ as a function of
%       $r_{\epsilon \ast}$ for different values of $\alpha_{\rm NFW}$, see
%       Eqs.~\eqref{eq:15b} and~\eqref{eq:20}. Notice that the critical
%       value corresponding to the soliton case, $\lambda_{\rm cr}
%      \simeq 0.48$, is obtained asymptotically in the limit
%       $r_{\epsilon \ast} \to \infty$. (Right) The same as before but
%       now for the total mass $M$ in Eq.~\eqref{eq:12}, with its value
%       normalized in terms of the soliton mass $M_s$ given in
%       Eq.~\eqref{}. The vertical black line in both plots marks the
%       position of the soliton radius, in dimensionless units, at
%       $r_{\epsilon \ast} = 1$. The values of the soliton profile
%       alone correspond to the limit $\alpha_{\rm NFW} \to \infty$. 
%      See the text for more details.}
\label{fig:Lambdacr}
       %Notice that the latter is the asymptotic value at large $r_{\epsilon \ast}$, whereas for small values of the latter the total mass $M$ can be three orders of magnitude larger than the soliton mass $M_s$. Here the upper limit of integration in Eq.~\eqref{eq:12} was taken as $r_\ast = 20$.
\end{figure*}

Going back to the complete profile (Eq.~\eqref{eq:profile}), we start with
the calculation of the critical value $\lambda_{\rm crit}$ from the
analytical formula in Eq.~\eqref{eq:15b}. The (total) projected surface mass 
density for the special value $\theta_\ast = 0$ is obtained from the
first branch, in Eq.~\eqref{eq:8a}, as
\begin{widetext}
  \begin{equation*}
    \Sigma_\ast(0,r_{\epsilon \ast},\alpha_{\rm NFW}) = 2 \left[ \int
    \limits^{r_{\epsilon \ast}}_0 \frac{dz}{\left( 1 + z^2 \right)^8}
    + \frac{r_{\epsilon \ast} \left( 1 + \alpha_{\rm NFW} \,
        r_{\epsilon \ast} \right)^2}{\left( 1+ r^2_{\epsilon \ast}
      \right)^8} \int \limits^\infty_{r_{\epsilon \ast}} \frac{dz}{z
      \left( 1 + \alpha_{\rm NFW} \, z \right)^2} \right] \,
  , \label{eq:20}
  \end{equation*}
\end{widetext}
which indicates, together with Eq.~\eqref{eq:11}, that the critical
value $\lambda_{\rm cr}$ of the combined profile (Eq.~\eqref{eq:profile}) is
a function of $r_{\epsilon \ast}$ and $\alpha_{\rm NFW}$, and its
behavior for different combinations of these parameters is shown in
the left panel of Fig.~\ref{fig:Lambdacr}.  {Notice that we have
       taken into account the constraint $r_{\epsilon \ast} \geq
       r_{\epsilon \ast, \rm max}$, see Eq.~\eqref{eq:16}. Moreover,
       it can be seen that the lowest value of $\lambda_{\rm cr}$, for
       any given value of $\alpha_{\rm NFW}$, is indeed attained at
       $r_{\epsilon \ast, \rm max}$ as indicated by the vertical lines with the corresponding colors}. Not surprisingly, the
addition of the NFW outer part helps the soliton profile to achieve small values of $\lambda_{\rm crit}$, which in turn eases the accomplishment of the inequality in Eq.~\eqref{eq:10}. In particular, Fig.~\ref{fig:Lambdacr} shows that $\lambda_{\rm crit} \to 0$ as $\alpha_{\rm NFW} \to 0$, which means that the combined profile~\eqref{eq:profile} will be able to produce a lensing signal for any non-trivial combination of its parameters $\rho_s$ and $r_s$.

In the case of the combined profile the total mass
$M(r)$ inside a sphere of any given radius $r > r_\epsilon$ is simply given by the integral
\begin{widetext}
\begin{equation}
  \label{eq:12}
  \frac{M(r)}{ 10^{13} M_\odot} = 3 \, \left( \frac{m_a}{10^{-22} \,
  \textrm{eV}} \right)^{-2} \left( \frac{r_s}{\textrm{pc}} \right)^{-1} \times \left[ \int \limits^{r_{\epsilon
        \ast}}_0 \frac{dx \, x^2}{(1 + x^2)^8} + \frac{r_{\epsilon
        \ast} \left( 1 + \alpha_{\rm NFW} \, r_{\epsilon \ast}
      \right)^2}{\left( 1+ r^2_{\epsilon \ast} \right)^8} \int
    \limits^{r_\ast}_{r_{\epsilon \ast}} \frac{dx \, x}{\left(1 +
        \alpha_{\rm NFW} \, x \right)^2} \right] \, .
\end{equation}
\end{widetext}
In the general case the total mass diverges as $r \to \infty$, whereas for the soliton profile only (which requires $r_{\epsilon \ast} \to
\infty$) we simply obtain that its total mass $M_s$ is \citep{Guzman:2004wj,marsh,Chen:2016unw}
\begin{equation}
  \label{eq:5}
  \frac{M_s}{ 10^{11} M_\odot} = 7.7 \, \left( \frac{m_a}{10^{-22} \,
  \textrm{eV}} \right)^{-2} \left( \frac{r_s}{\textrm{pc}}
\right)^{-1} \, .
\end{equation}

In general, we expect from Eq.~\eqref{eq:12} the total mass in the
combined profile to be larger than the soliton alone, that is $M(r)
\geq M_s$. However, the value of the total mass $M$ will depend on the
upper limit of integration $r_\ast$, and the largest values for any
given $r_\ast$ will be obtained for the case where $\alpha_{\rm NFW} \to 0$, similar to the case of the critical value $\lambda_{\rm crit}$. The
aforementioned general behaviour of the total mass $M$ as a function of
the free parameters $r_{\epsilon \ast}$ and $\alpha_{\rm NFW}$ is
shown in the right-hand panel of Fig.~\ref{fig:Lambdacr}. For the numerical examples we considered the upper limit of integration $r_\ast =20$, for which we then see that that the difference between $M$ and $M_s$ can be as large as three orders of magnitude in the case $\alpha_{\rm NFW} =0$.  {In other words, the total mass in the WaveDM profile is intrinsically attached to that of its soliton, and then the latter should be large enough if we are going to get the right mass scales in galaxies. This is a non-trivial property, as it shows that any non-zero value of the parameter $\alpha_{\rm NFW}$ could point out the existence of a soliton core with a non-negligible mass contribution to the lens system (see for instance Fig.~(\ref{fig:0004}) below)}.

\section{\label{sec:dat}Data analysis}

In this section we will  use our theoretical results to infer
information about the WaveDM profile from observations of specific
lens systems. We recall from Sec.~\ref{sec:den} that there
are four free parameters that are needed to describe the lensing
properties of the combined density profile,
Eq.~\eqref{eq:profile}. However, the lens equation,
discussed in Sec.~\ref{sec:grav}, is not explicitly dependent on two
of them, namely $\rho_s$ and $r_s$, and depends only on the
free parameters of the NFW outer profile $r_{\epsilon \ast}$ and $\alpha_{NFW}$. Therefore we could use the right-hand side of the lens equation (Eq.~\eqref{eq:19a}) to put constraints on the surface density through the  combination of parameters $\rho_s r_s$ -- see also the discussion in Sec.~\ref{sec:lensing-equations-}.

However, the special properties of the WaveDM profile, as
represented by Eq.~\eqref{eq:2}, suggest that the lens equation could 
be written in a more convenient form. Using the fact that the (normalized) angular Einstein radius is $\theta_{\ast E} = R_E/r_s$, Eq.~\eqref{eq:lenseq} can be re-cast in the form
\begin{equation}
   m^{-2}_{a22} \, \theta_{\ast E} \, m_\ast(\theta_{\ast E},\alpha_{\rm
     NFW}, r_{\epsilon \ast}) = \frac{1}{2.4}
   \frac{d_{OS}}{d_{OL} d_{LS}} \frac{h}{0.57} \left( \frac{R_E}{\rm
       kpc} \right)^3 \, , \label{eq:24}
\end{equation}
where we have set $m_{a 22} \equiv m_a/10^{-22} {\rm
  eV}$. Equation~\eqref{eq:24} then defines a different observable,
which results solely from the combination of the distances involved in
the measurement of the lens system, so that we can put constraints
directly on the boson mass $m_a$ rather than on the energy density
$\rho_s$, but in any case in combination with the rest of parameters,
namely $\theta_{\ast E}$, $\alpha_{\rm NFW}$, and $r_{\epsilon \ast}$.

\begin{table*}[htp!]
	\begin{center}
		\renewcommand{\arraystretch}{2.0}
		\begin{tabular}{|l | c | c | c | c
                  | c|}
			\hline
			Name  & $f^{Salp}_{\ast, \rm{Ein}}$ &
                                                                     $z_{ \rm lens}$ & $z_{ \rm source}$ & $d_{OS}/(d_{OL} d_{LS})$ & $R_E$ (kpc)  \\ \hline
                SLACS  & \multicolumn{5}{c|}{} \\
                \hline
			J0008-0004 & $0.50 \pm 0.16$  & 0.44 & 1.192 & 6.6855 & 6.7965  \\ 
			J0935-0003 & $0.35 \pm 0.05$ & 0.347 & 0.467 & 18.2172 & 4.4063  \\  
			J0946+1006 & $0.46 \pm 0.13$ & 0.222 & 0.609 & 9.7613 & 5.0934  \\ 
			J1143-0144 & $0.46 \pm 0.10$ & 0.106 & 0.402 & 14.9617 & 3.3683 \\
			J1306+0600 & $0.47 \pm 0.08$ & 0.173 & 0.472 & 11.7208 & 4.0050  \\
			J1318-0313 & $0.42 \pm 0.08$ & 0.24 & 1.3 & 7.2634 & 6.1840 \\ \hline 
			LSD  & \multicolumn{5}{c|}{} \\ \hline
			CFRS03.1077 & $0.46 \pm 0.15$  & 0.94 & 2.94 & 5.3188 & 10.0470  \\ 
			HST1417+5226 & $0.38 \pm 0.11$  & 0.81 & 3.40 & 4.7801 & 10.9360  \\ 
			\hline
			SL2S  & \multicolumn{5}{c|}{} \\ \hline
			J220329+020518 & $0.24 \pm 0.06$  & 0.40 & 2.150 & 5.4526 & 10.8130  \\ 
			\hline
\end{tabular}
\caption{ {Selected galaxies from SLACS, LSD and SL2S. Columns correspond to: label within the SDSS catalog (Name), fraction of luminous matter within the Einstein radius ($f^{Salp}_{\ast, \rm{Ein}}$), redshift of the lens ($z_{ \rm lens}$) and the source ($z_{ \rm source}$), distance factor $d_{OS}/(d_{OL} d_{LS})$, and measured Einstein radius ($R_E$). Selection was based on  the condition $ f^{Salp}_{\ast, \rm{Ein}} \leq 0.5$. } }
%\caption{List of selected galaxies from SLACS, LSD and SL2S. These were selected because they have a fraction of luminous matter of $0.5$ or less; see the values in the second column. Column (1) gives the label of the galaxies within the SDSS catalog, column (2) indicates the fraction of luminous matter. Column (6) lists the measured Einstein radius in units of $\textrm{kpc}$. } %{\bf Explain here the error propagation in the ratios of the reduced distances and adjust the significant figures accordingly.}
		\label{tab:4}
	\end{center}
\end{table*}

% \begin{table*}[htp!]
% 	\begin{center}
% 		\renewcommand{\arraystretch}{2.0}
% 		\begin{tabular}{l | c | c | c | c
%                   | c}
% 			\hline
% 			Name  & $f^{Salp}_{\ast, \rm{Ein}}$ &
%                                                                      $z_{ \rm lens}$ & $z_{ \rm source}$ & $d_{OS}/(d_{OL} d_{LS})$ & $R_E$  \\ \hline
% 			J0008-0004 & $0.50 \pm 0.16$  & 0.44 & 1.192 & 6.609565 & 6.59 \\ 
% 			J0935-0003 & $0.35 \pm 0.05$ & 0.347 & 0.467 & 18.04391 & 4.26 \\  
% 			J0946+1006 & $0.46 \pm 0.13$ & 0.222 & 0.609 & 9.700301 & 4.95 \\ 
% 			J1143-0144 & $0.46 \pm 0.10$ & 0.106 & 0.402 & 14.9161 & 3.27 \\
% 			J1306+0600 & $0.47 \pm 0.08$ & 0.173 & 0.472 & 11.66306& 3.87 \\
% 			J1318-0313 & $0.42 \pm 0.08$ & 0.24 & 1.3 & 7.215974 & 6.01 \\ \hline  
% \end{tabular}
% \caption{List of selected galaxies from SLACS. These were selected because they have a fraction of luminous matter of $0.5$ or less; see the values in the second column. Column (1) gives the label of the galaxies within the SDSS catalog, column (2) indicates the fraction of luminous matter. Column (6) lists the measured Einstein radius in units of $\textrm{kpc}$. {\bf Explain here the error propagation in the ratios of the reduced distances and adjust the significant figures accordingly.}}
% 		\label{tab:4}
% 	\end{center}
% \end{table*}

In general, we expect that, given the data from a single galaxy, there
will always be a region in the parameter space that will satisfy Eq.~\eqref{eq:24}. Thus, for a given sample of galaxies, we could in principle determine the range of possible values of $m_a$ that is consistent with the observed data. However, we must recall
that the boson mass $m_a$ is a fundamental physical parameter of the
model which in principle should have a unique value. This means that
the boson mass should be treated differently from other parameters in the
model and should {\em not} be given the freedom to vary from galaxy to galaxy. 

Our proposal, therefore, is to study the lensing properties of the WaveDM profile by
fixing the value of the boson mass and finding, via statistical analysis, the best-fit values of the remaining free parameters $\theta_{\ast E}$,
$\alpha_{\rm NFW}$ and $r_{\epsilon \ast}$.
 {As we are interested in
the properties of the WaveDM profile alone, 
we select a particular sub-sample of
early-type galaxies, we will focus in particular on those lensed systems in which the galaxy is known to have a relatively high dark matter fraction. We have set a threshold at a fraction of luminous matter of $50$\% or less, i.e. reducing as much as possible its effects. Also, due to the consideration of the lens model to be comprised of a single galaxy, samples where contribution of more than one component is known were excluded, e.g. MG 2016+112~\cite[see][]{MG2016}.}

 {We use strongly-lensed galactic-scale systems observed by the Sloan data from the Sloan Lens ACS (SLAC) survey, which is comprised of nearly 100 likely and confirmed lensed systems~\cite[see][]{SlacsIV,SlacsIX}. Our criteria reduces this sample to a sub-sample of only 6 galaxies.}
 {We also include samples from Lens Structure and Dynamics(LSD), and Strong Lensing Legacy Survey(LS2S)~\cite[see e.g.][]{cao2015,sL2SIII,LSD}, but after applying the same criteria we ended selecting only three galaxies. The names of the nine chosen galaxies are shown in Table~\ref{tab:4} together with the values of their lens parameters.}

 {Another advantage from the SLAC survey is it was analized previously with other, similar, scalar field dark matter models~\cite[see e.g.][]{GonzalezMorales:2012uw,robles2013}. Due to the lack of studies of lensing for this kind of model, we therefore adopt SLACS as the main sub-sample for consistency with previous studies and as a proof of concept for the possible use of the methodology.}

The Einstein radius
$R_E$ is obtained using $R_E = D_{OL}\theta_E$, where $\theta_E$ is as given in~\cite[see][]{SlacsV,SlacsIX,SL2SIV,LSD}.
 {It has been shown that $\theta_E$ is nearly model independent and well constrained, and have been used before to determine the lens mass where only big asymmetric arcs in the images could produce some difference\footnote{In particular for SLACS, it is important to remark that
they adjusted different mass models and found that the images where visually indistinguishable and the Einstein angles where the same within error. This is stated in section section 5.2 and Table 5 of \cite{SlacsV}.}~\cite[see e.g.][]{Kochanek91,kochanek2001,trick2016,tortora2018,lyskova2018,cardone2009}}; this justifies its direct use as a reliable observable in our analysis. The 
distances and $R_E$ values in Table~\ref{tab:4} are obtained considering a cosmology with matter-density parameter $\Omega_M = 0.3089 $, vacuum energy-density parameter $\Omega_\Lambda = 0.6911$, and Hubble parameter $H_0 = 67.74\,  \mbox{km}\, \mbox{s}^{-1} \, \mbox{Mpc}^{-1}$ from the ~\cite{Planckresults}. 

% Antonio: I found that it does not matter that the Einstein radius 
%was adjusted by  an SIE or SIS. It should be considered model independent and this will suffice.  

\subsection{Soliton core profile}
As a first case of study, let us consider the soliton core profile
without the external NFW part. There are in this case only two free
parameters: $m_{a22}$ and $\theta_{\ast E}$. In section~\ref{sec:complete} a
Bayesian analysis will be carried out, taking into account the results from this section. 
The projected mass surface density given by Eq.~\eqref{eq:14}, with the help of 
Eq.~\eqref{eq:14a}, has in this case an analytical expression,
\begin{equation}
  \label{eq:17}
  m_\ast(\theta_{\ast E}) = \frac{2}{13 \lambda_{\rm crit}} \,
  \frac{(1 + \theta^2_{\ast E})^{13/2} -1}{(1 + \theta^2_{\ast
      E})^{13/2}} \, ,
\end{equation}
where $\lambda_{\rm crit} \simeq 0.484$ is the critical value
calculated from Eq.~\eqref{eq:15b}; see also Table~\ref{tab:1}. Notice
that $m_\ast(0) = 0$, whereas its asymptotic limit is $m_\ast(\infty)
= 2/(13 \lambda_{\rm crit})$.

\begin{figure}[htp!]
  \centering
    \includegraphics[width=0.49\textwidth]{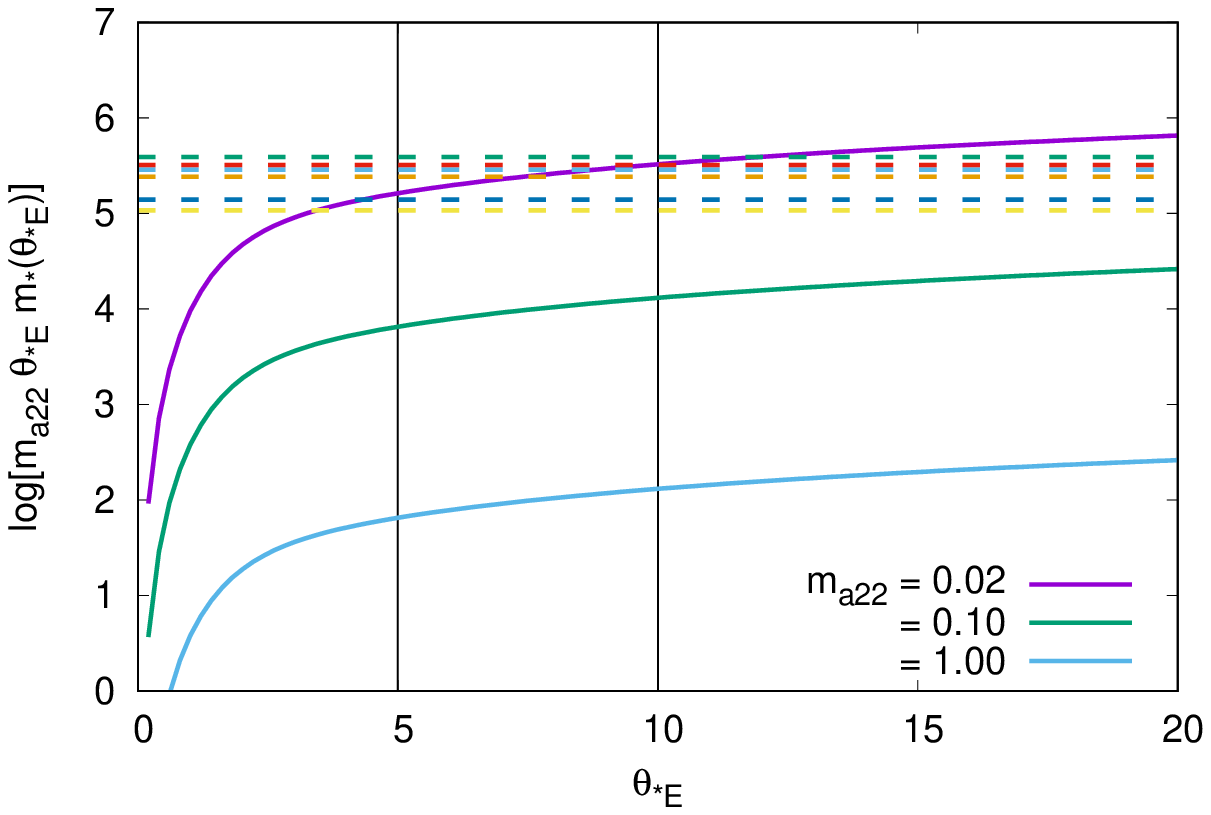}
    \includegraphics[width=0.49\textwidth]{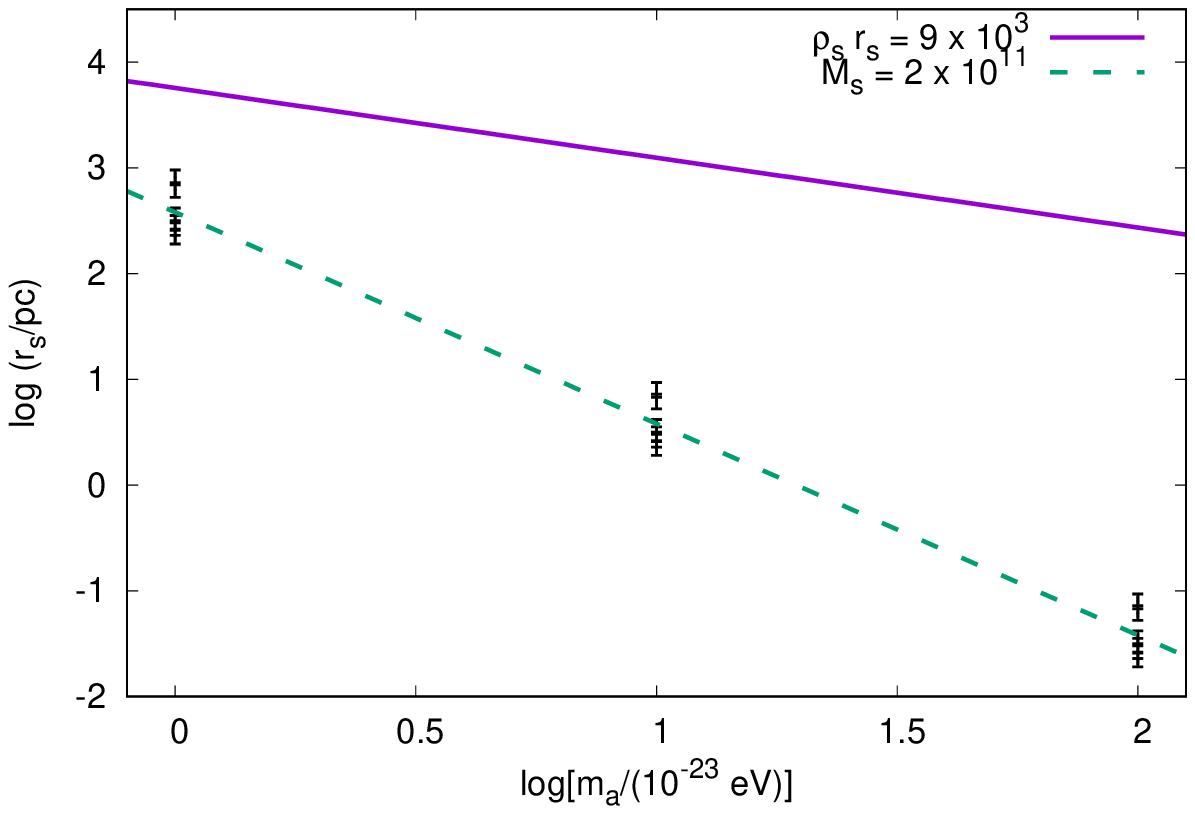}
    \caption{ {(Left) Left hand side of  Eq.~\eqref{eq:24} as function of the normalized Einstein angle, $\theta_{\ast E} = R_E/r_s$, according to Eq.~\eqref{eq:17}. The intersection with the dashed horizontal lines, r.h.s of Eq.~\eqref{eq:24} for each galaxy in our sample defines the value of $\theta_{\ast E}$, i.e the scale radius. (Right) Fitted soliton radius, $r_s$, as a function of  the boson mass $m_a$. The resultant $r_s$ lies along the line of constant soliton mass $M_s$. This example is for $M_s \simeq 2 \times 10^{11} \, M_\odot$ . See the text for more details. }}
%     \caption{(Left) Illustration of how we can use  to     constrain the parameters of the WaveDM model.  The curves show,   for selected values of the boson mass $m_{a 22}$, how the left-hand side of Eq.~\eqref{eq:24} varies as a function of the Einstein angle $.  using the right-hand side of the aforementioned equation. (Right) The best fit values of the soliton radius $r_s$ shown in Table~\ref{tab:5} for fixed values of the boson mass$m_a$. Because the main constraint imposed by the lensing system is for the total mass inside the Einstein radius $R_E$, the obtained data points lie along the line of constant soliton mass $M_s \simeq 2 \times 10^{11} \, M_\odot$. See the text for more details.}
       %From these examples, it can be seen that the preferred values of the boson mass appear to be $m_{a22} \simeq 0.02$.
       %The data points also lie below the line representing the inequality~\eqref{eq:10} for the surface density $\rho_s \, r_s = 9 \times 10^3 \, M_\odot \, \textrm{pc}^{-2}$
\label{fig:lensing}
\end{figure}

To obtain a basic understanding of the solutions that will be found for
the physical parameters, we show in the left panel of
Fig.~\ref{fig:lensing} the expected behavior of the left-hand side of
Eq.~\eqref{eq:24} as a function of the Einstein angle $\theta_{\ast
  E}$. We also show, as the series of horizontal lines, the values of 
  the right-hand side of Eq.~\eqref{eq:24} obtained from the 
  observed data for the galaxies listed in
Table~\ref{tab:4}.

Figure~\ref{fig:lensing} shows that it will always be possible to identify a value of the Einstein angle $\theta_{\ast E}$ for which the left-hand and right-hand sides of Eq.~\eqref{eq:24} are in agreement, irrespective of the value of
the boson mass -- although as the boson mass increases the agreement occurs at increasingly large values of $\theta_{\ast E}$. For the examples shown in Fig.~\ref{fig:lensing}, a
boson mass of order $m_{a22} \simeq 0.02$ seems to fit well the
SLACS galaxies listed in Table~\ref{tab:4} -- corresponding to an allowed range 
for the angular Einstein radius of $5 < \theta_{\ast E} < 10$. This latter
range can also be translated into an allowed range for the soliton
radius, and suggests that $r_s \sim {\rm kpc}$ for the given example galaxies. However, note that it is always possible to find a solution that matches the left-hand and right-hand side of Eq.~\eqref{eq:24}, for any given value of the boson mass $m_a$, by a suitably large choice of Einstein angle $\theta_{\ast E}$, that is, by choosing $r_s \to 0$. We must recall that the latter condition means that the density profile is dominated by a very massive and compact soliton, but this can be in disagreement with other indications about the actual size of the dark matter halo in the lens galaxies.

To summarize, given that we have only one observable constraint, the 
most we can do is first to fix the value of the boson mass $m_a$ and 
from this to obtain constraints on the remaining free parameters that 
are consistent with that boson mass. Specifically, by adopting a proposed  value for the boson mass $m_a$ in Eqs.~\eqref{eq:24} and~\eqref{eq:17},  we can obtain for each galaxy the corresponding best-fit value for  $\theta_{\ast E}$, and from that the best-fit value for $r_s$. 
%\textcolor{blue}{Regarding the luminous matter, it was found in our tests that its inclusion made no difference in the resultant values. We shall explain in the next section how we introduced it in our analysis.}
%Antonio: I run the tests, and it does not make too much difference by the technique
%we are using.

The results obtained for our selected sample of galaxies are shown in
Table~\ref{tab:5}, and also plotted in the right panel of
Fig.~\ref{fig:lensing}. The latter figure speaks for itself, and shows that the data points for all galaxies lie along the line with a 
constant soliton mass $M_s \simeq 10^{11} \, M_\odot$ 
(see Eq.~\eqref{eq:5}), and (as required) all lie below the line that 
represents the inequality, Eq.~\eqref{eq:10}, for the
galaxy in Table~\ref{tab:4} (J0935-0003) with the most extreme 
value for the ratio of distances on the right-hand side of Eq.~\eqref{eq:24}.
The different values obtained for the characteristic radius $r_s$ give
an enclosed mass which corresponds closely to the values reported in
\cite{SlacsIX}. Nevertheless, these models are found to be considerably 
too compact when the characteristic radius and corresponding enclosed mass 
are considered together.  For example, galaxy J0008-0004 has a value for 
$M_{\rm Eins} = 3.1\times10^{11} \, M_\odot$ which is comparable with 
the value of $M_s =  3.4\times10^{11} \, M_\odot$ obtained using the best 
fit parameters of the soliton model. Notwithstanding that the soliton model 
gives an enclosed mass that is adequate and realistic, we think that the
characteristic radius is most definitely not so. This is by taking into 
consideration the results from rotation curves where the effects of
dark matter are expected to be at larger radii than the luminous part
of the galaxy, and this contrasts with the values obtained for the soliton
alone where the mean effective
radius for galaxy J0008-0004 is observed to be $r_e \approx 9.6 \,
\textrm{kpc}$, which is several orders of magnitude larger than 
the characteristic radius $r_s$ obtained for any of the different
boson masses presented in Table~\ref{tab:5}, including the samples from the other surveys.  Therefore we think the
soliton profile alone is actually not helping to explain the
distribution of dark matter around the selected galaxies in a
consistent way.

\begin{table}[htp!]
	\begin{center}
		\renewcommand{\arraystretch}{2.0}
		\begin{tabular}{| l | c | c | c |}
			\hline
	& $m_{a22} =10$ & $m_{a22} = 1 $ & $ m_{a22} = 0.1 $ \\
	\hline
	Galaxy  & \multicolumn{3}{c|}{$\log_{10} (r_s/\textrm{pc})$} \\
	 \hline
	SLACS  & \multicolumn{3}{c|}{} \\
	\hline
	J0008-0004 &  $-1.67^{+0.07}_{-0.06}$ & $0.33^{+0.07}_{-0.06}$ & $2.33^{+0.07}_{-0.06}$ \\ 
	J0935-0003 & $-1.73^{+0.07}_{-0.06}$ & $0.27^{+0.07}_{-0.06}$ & $2.27^{+0.07}_{-0.06}$ \\  
	J0946+1006 & $-1.59^{+0.07}_{-0.06}$ & $0.41^{+0.07}_{-0.06}$ & $2.41^{+0.07}_{-0.06}$ \\ 
	J1143-0144 & $-1.41^{+0.07}_{-0.06}$ & $0.59^{+0.07}_{-0.06}$ & $2.59^{+0.07}_{-0.06}$ \\
	J1306+0600 & $-1.46^{+0.07}_{-0.06}$ & $0.54^{+0.07}_{-0.06}$ & $2.54^{+0.07}_{-0.06}$ \\
	J1318-0313 &$-1.62^{+0.07}_{-0.06}$ & $0.37^{+0.07}_{-0.06}$ & $2.37^{+0.07}_{-0.06}$ \\ \hline 
	LSD  & \multicolumn{3}{c|}{} \\
	\hline
	CFRS03.1077 &  $-1.58^{+0.17}_{-0.12}$ & $0.42^{+0.17}_{-0.12}$ & $2.42^{+0.17}_{-0.12}$ \\ 
	HST1417+5226 &  $-1.7^{+0.18}_{-0.12}$ & $0.30^{+0.18}_{-0.12}$ & $2.30^{+0.17}_{-0.12}$ \\ 
	\hline
	SL2S  & \multicolumn{3}{c|}{} \\
	\hline
	J220329+020518 &  $-1.90^{+0.06}_{-0.05}$ & $0.10^{+0.06}_{-0.05}$ & $2.10^{+0.06}_{-0.05}$ \\ \hline
	\end{tabular}
	
%	\caption{The values of the soliton radius in the logarithmic scale $\log_{10} (r_s/\textrm{pc})$ obtained from the fits to the indicated galaxies, for three different values of the boson mass $m_a$. The data points for the SLACS data are also shown in the bottom panel of Fig.~\ref{fig:lensing}, where we see that they all indicate that the total mass contained within the Einstein radius is $M_s \simeq 10^{11.5} \, M_\odot$. The LSD and SL2S samples have a total mass of } $M_s \simeq 10^{11.8}\, M_\odot$
          %{\bf I updated the values considering Planck parameters. They changed slightly, but I can explain this from my side.}}
	\end{center}
	\caption{ { Soliton radius, $r_s$, obtained from the fits to each galaxy and for three different values of the boson mass $m_a$. Note that for SLACS samples, all combinations have a total soliton mass mass contained within the Einstein radius of $M_s \simeq 10^{11.5} \, M_\odot$. For the LSD and SL2S samples have a total mass of $M_s \simeq 10^{11.8}\, M_\odot$ } }
	\label{tab:5}

\end{table}

% \begin{table}[htp!]
% 	\begin{center}
% 		\renewcommand{\arraystretch}{2.0}
% 		\begin{tabular}{ l | c | c | c }
% 			\hline
% 	& $m_{a22} =10$ & $m_{a22} = 1 $ & $ m_{a22} = 0.1 $ \\
% 	\hline
% 	Galaxy  & \multicolumn{3}{c}{$\log_{10} (r_s/\textrm{pc})$} \\
% 	 \hline
% 	J0008-0004 &  $-1.65^{+0.07}_{-0.06}$ & $0.35^{+0.07}_{-0.06}$ & $2.35^{+0.07}_{-0.06}$ \\ 
% 	J0935-0003 & $-1.52^{+0.07}_{-0.06}$ & $0.48^{+0.07}_{-0.06}$ & $2.48^{+0.07}_{-0.06}$ \\  
% 	J0946+1006 & $-1.45^{+0.07}_{-0.06}$ & $0.55^{+0.07}_{-0.06}$ & $2.55^{+0.07}_{-0.06}$ \\ 
% 	J1143-0144 & $-1.10^{+0.07}_{-0.06}$ & $0.90^{+0.07}_{-0.06}$ & $2.91^{+0.07}_{-0.06}$ \\
% 	J1306+0600 & $-1.21^{+0.07}_{-0.06}$ & $0.79^{+0.07}_{-0.06}$ & $2.79^{+0.07}_{-0.06}$ \\
% 	J1318-0313 &$-1.57^{+0.07}_{-0.06}$ & $0.43^{+0.07}_{-0.06}$ & $2.43^{+0.07}_{-0.06}$ \\ \hline  
% 	\end{tabular}
% 	\caption{The values of the soliton radius in the
%           logarithmic scale $\log_{10} (r_s/\textrm{pc})$ obtained from the
%           fits to the indicated galaxies, for three different values
%           of the boson mass $m_a$. The data points are also shown in
%           the bottom panel of Fig.~\ref{fig:lensing}, where we see
%           that they all indicate that the total mass contained within
%           the Einstein radius is $M_s \simeq 10^{11} \,
%           M_\odot$.}
% 	\label{tab:5}
% 	\end{center}
% \end{table}

There are two valuable lessons from the above exercise. The first one
is that the soliton core profile alone will always be able to fulfill
the lensing constraints even without the consideration of the NFW
contribution given the Einstein-radius as the only measurement to
satisfy. This is not surprising, as the lensing equations can be
solved even if we consider a point particle with the required total
mass (which formally corresponds to the soliton core profile with $m_a
\to \infty$). The second lesson is that even though the soliton
profile may be adequate, formally speaking, to explain the lensing
properties of the galaxies in Table~\ref{tab:4}, we will, in any case, 
have to consider the NFW outskirts in the complete 
profile (Eq.~\eqref{eq:profile}) in order to satisfy other constraints that 
suggest that the boson mass should be in the range 
$m_{a22} = 1-10$~\cite[see][]{Hui:2016ltb}.

\subsection{\label{sec:complete} Complete profile}
Taking into account the above experience gained with the soliton profile
alone, we will now consider the following procedure for the complete
WaveDM profile. Since the total mass inside the Einstein radius is the
only constraint provided by the lens systems, we will fix the values
of the boson mass $m_a$ and soliton mass $M_s$. This approach is considered
due to the results from the soliton analysis where the boson mass can satisfy
different values for the Einstein radius, and other studies have found that $m_a$ 
needs to be in certain range. For this, we take
following values of the boson mass $m_{a22} = 0.1,1,10$, and for the
soliton mass $\log_{10}(M_s/M_\odot) = 11.5,10.5,9.5,8.5,7.5$, from
which we will calculate the values of $r_s$ by means of
Eq.~\eqref{eq:5}, which allows to avoid a possible overcompensation of the soliton mass.

We will adopt a uniform prior for the other parameters over the
following ranges: $\alpha_{\rm NFW} = [0:10]$, and $r_{\epsilon
\star} = [r_{\epsilon \star, max}:10]$. Here $r_{\epsilon \ast, \rm
max}$ is found from the cubic equation~\eqref{eq:16} for a given
value of $\alpha_{\rm NFW}$, and the extreme values $\alpha_{\rm NFW}
= 10$ and $r_{\epsilon \ast} =10$ are suggested by
Figs.~\ref{fig:Lambdacr} and~\ref{fig:lensing}.

We will obtain the values of $\theta_{\ast E}$ by sampling from 
a Gaussian distribution, using the relation 
\begin{equation}
\theta_{\ast E}(p) = \theta_{\ast Em}+\sigma\sqrt{2}{\rm erf}^{-1}(2p-1)\,,\quad p\in(0,1). \label{eq:gauss}
\end{equation}  
The value for $\theta_{\ast Em} = R_E/r_s$ is the mean of the distribution using
the observed value for the Einstein radius, $\sigma = 0.05*\chi$ 
the error assigned and $p$ is a random number sampled from a uniform distribution on the interval [0,1]. The inverse error function is 
approximated as described in \cite{erf}. In this way, $\theta_{\ast E}$ will not enter into the fitting analysis as an extra variable.

Once the soliton mass is fixed, the rest of the mass that is included 
within the Einstein radius must be completed by the NFW
profile. Because this requires a huge contribution, up to three orders
of magnitude more, one sensible consideration is to set the total mass
of the lens as composed by a simple representation of luminous matter and
the selected model of dark matter. 
%to include a simple 
%approximation of a partial contribution of the luminous matter inside
%the Einstein radius. 
In a first approximation, the mass corresponding
to the baryonic matter is simply a constant value modeled as a point
particle. This  is done from Eq.~\eqref{eq:10c}, and then the
projected mass for the lens is composed of two parts,
\begin{equation}
m'(\theta) = m(\theta)+ M', 
\end{equation}
where $m(\theta)$ is the mass from the dark matter component given by
the profile in Eq.~\eqref{eq:profile}, and $M'= f{*,_{Ein}}M_{Ein}$
is the stellar mass contribution as described in
Table~\ref{tab:4}. These  values are normalized accordingly and then
the dimensionless total mass $m^\prime$ is
\begin{subequations}
\begin{equation}
m'_*(\theta_{\ast E},\alpha_{NFW},r_{\epsilon\ast}) = 
m_*(\theta_{\ast E},\alpha_{NFW},r_{\epsilon\ast}) +
M'_\ast \, , \label{eq:7}
\end{equation}
where
\begin{equation}
M'_\ast =0.3208 f_{\ast,Ein} \left(\frac{M_{Ein}}{M_s}\right).
\end{equation}
\end{subequations}
Eq.~\eqref{eq:7} is combined with Eq.~\eqref{eq:24} to produce a modified
observable which uses the soliton mass directly,
\begin{equation}
   \frac{M_s}{M_\odot} \, m^\prime_\ast(\theta_{\ast E},\alpha_{\rm
     NFW}, r_{\epsilon \ast}) = \frac{7.7\times10^8}{2.4}
   \frac{d_{OS}}{d_{OL} d_{LS}} \frac{h}{0.57} \left( \frac{R_E}{\rm
       kpc} \right)^2 \, . \label{eq:24mod}
\end{equation}

\subsection{General results}
Using the samples mentioned in section~\ref{sec:dat} we will try
to constrain the free parameters that will satisfy
Eq.~\eqref{eq:24mod}. As said before, the information available from
the data is the Einstein radius, $R_E$, the lens distances ($ d_{\rm
  OL}$, $d_{\rm LS}$, $d_{\rm OS}$), the lens redshift and the source redshift. This information is used in the Multinest code~\citep{multinest}
to carry out a parameter search for each individual galaxy.  { We carried out the analysis on the nine galaxies of our sub-sample of the surveys.} Nevertheless, they showed a similar behaviour for the range of values of $f_*$. 
\begin{figure*}[htp!]
  \centering
     \includegraphics[width=0.49\textwidth]{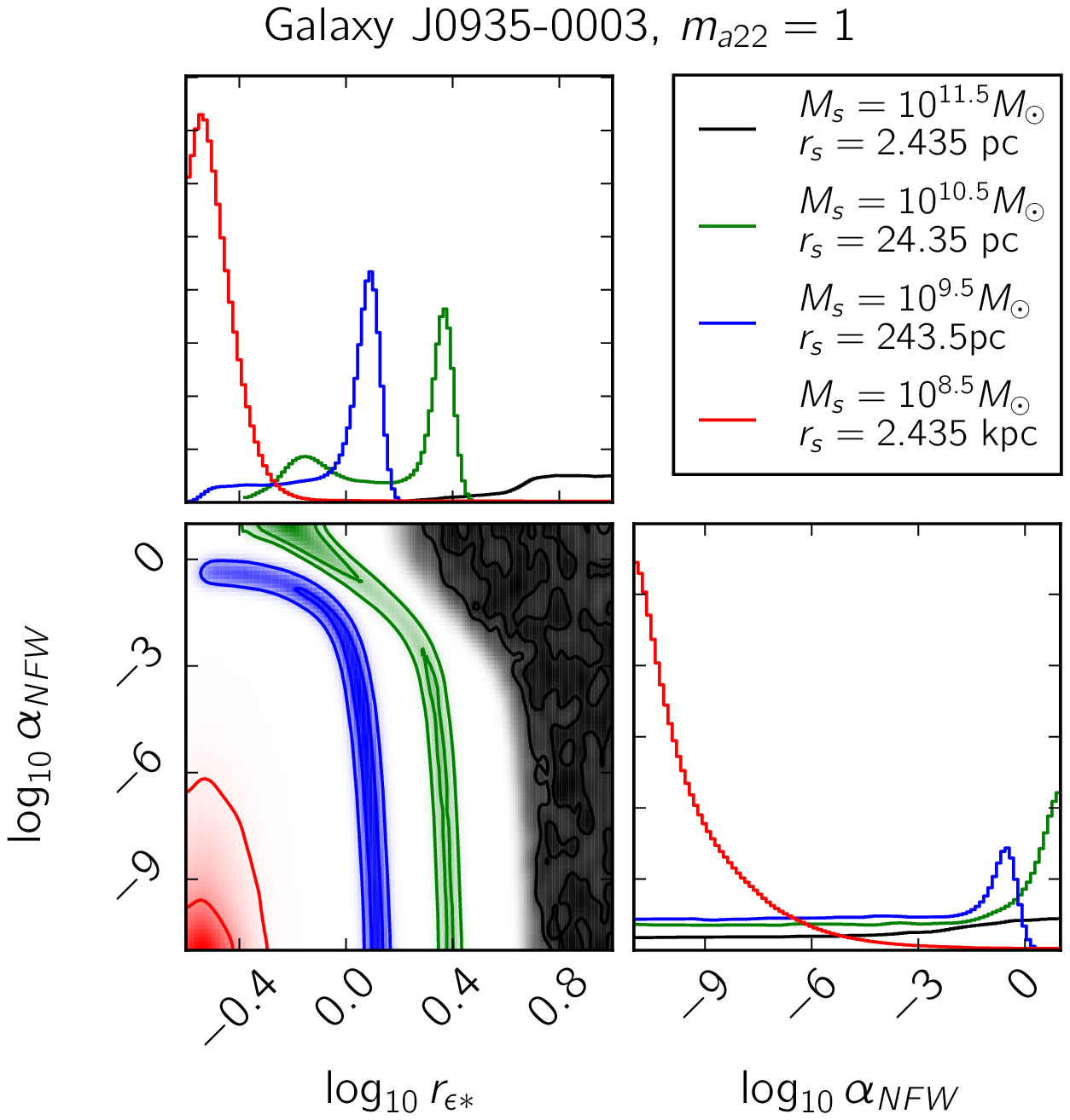}
    \includegraphics[width=0.49\textwidth]{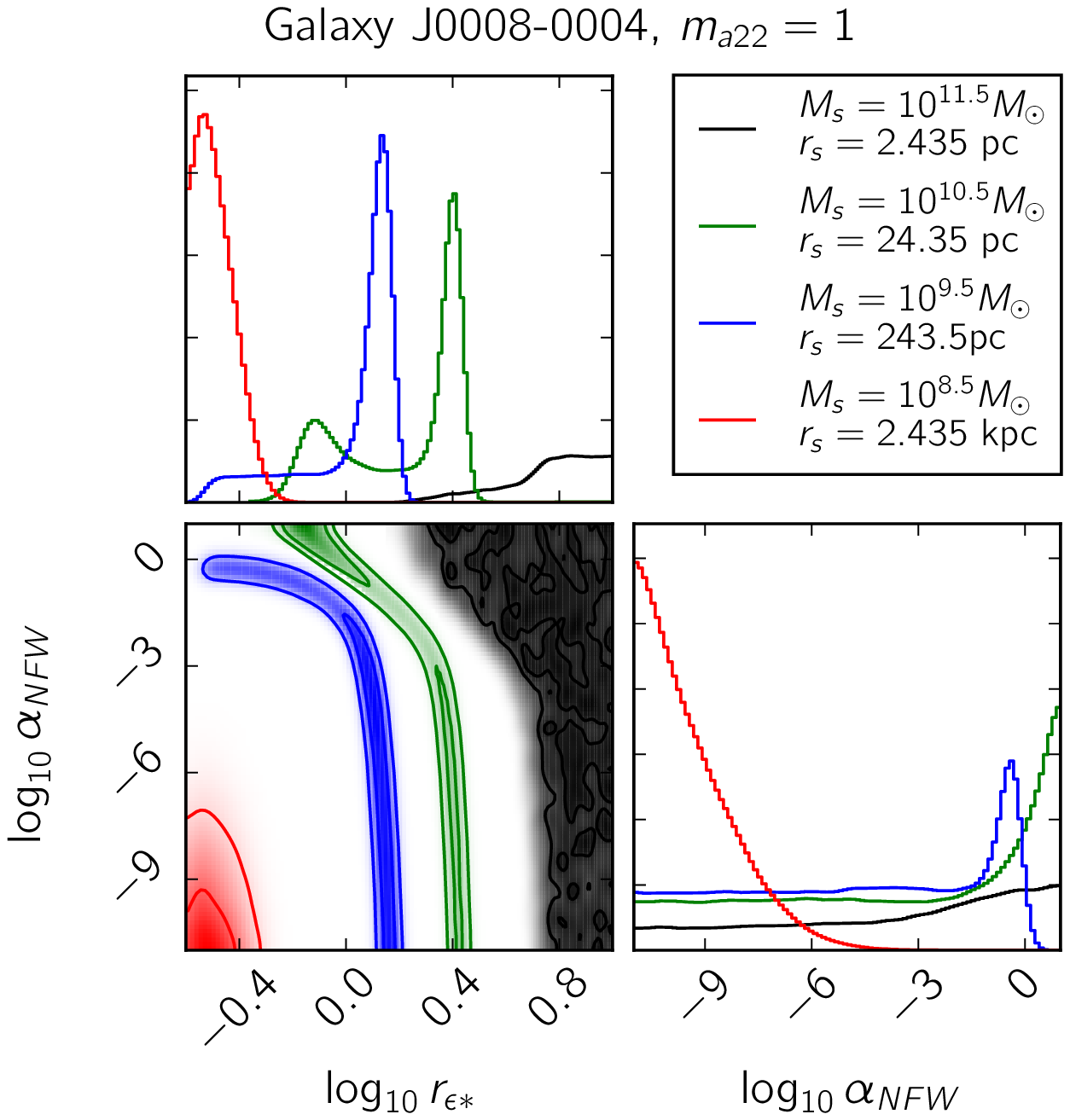}
    \includegraphics[width=0.49\textwidth]{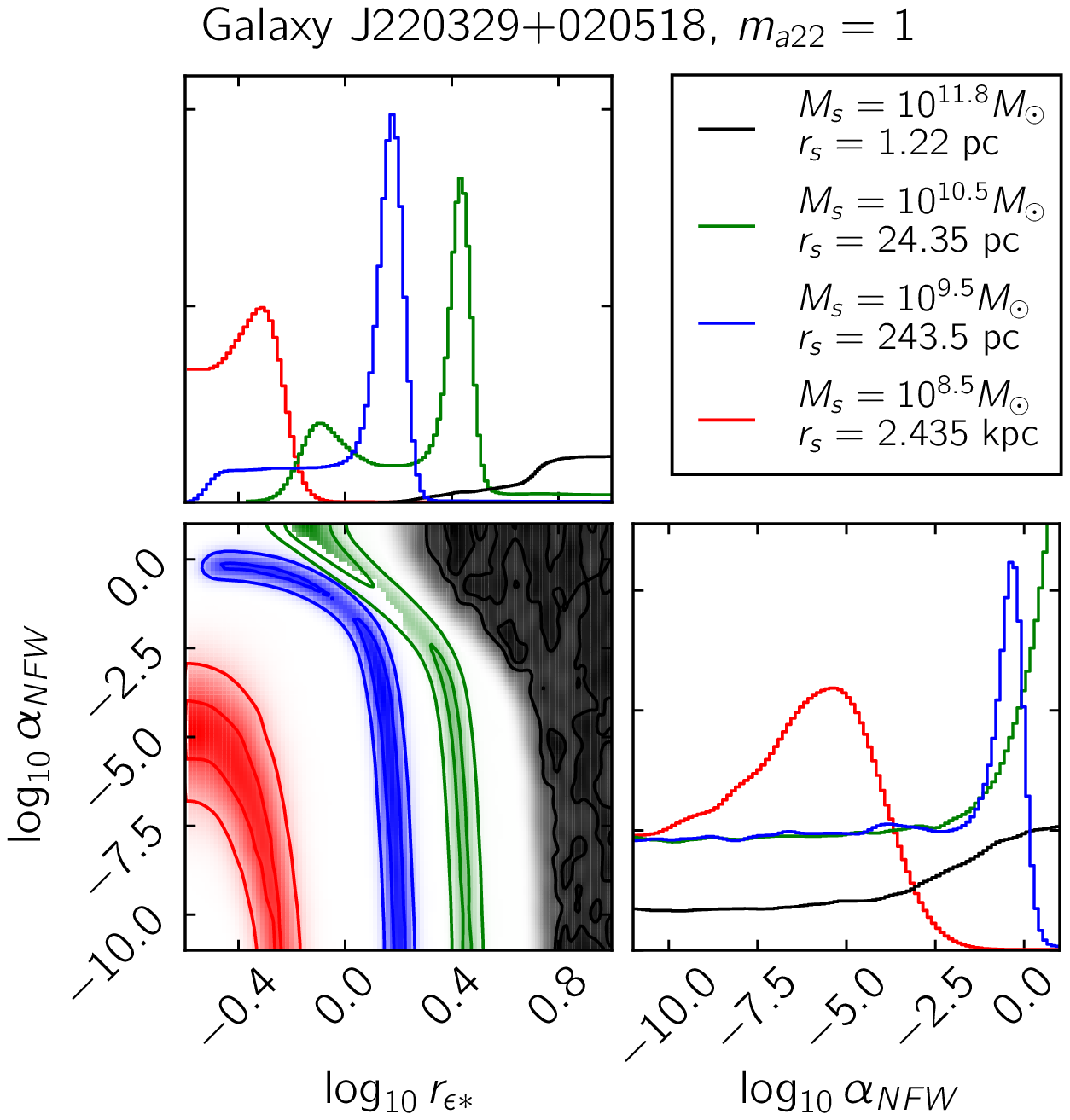}
        \caption{Posterior distribution for  the parameters fitted to galaxies
       {J0935-0003 (left), J0008-0004 (right) and J220329+020518(bottom); the contribution of
      the luminous matter are $35\%$, $50\%$ and $24\%$, respectively, of the total reduced mass
      inside the Einstein radius}.%, see also Eq.~\eqref{eq:24mod}. 
      The colors indicate different combunations of the soliton mass, $M_s$, and scale radius, $r_s$, computed with a fixed (normalized) boson mass $m_{a22}=  1$.}
%    \caption{Posterior distribution for  the parameters fitted to galaxies  {J0935-0003 (left), J0008-0004 (right) and J220329+020518(bottom); the contribution of the luminous matter are $35\%$, $50\%$ and $24\%$ of the total reduced mass  the Einstein radius}, see also Eq.~\eqref{eq:24mod}. The colors indicate different choices for the soliton mass $M_s$, and the values of the corresponding $r_s$, calculated from a fixed (normalized) boson mass $m_{a22}  =  1$, are also shown for comparison.}
    \label{fig:0004}
  \end{figure*}

  \begin{figure}[htp!]
  \centering
    \includegraphics[width=0.49\textwidth]{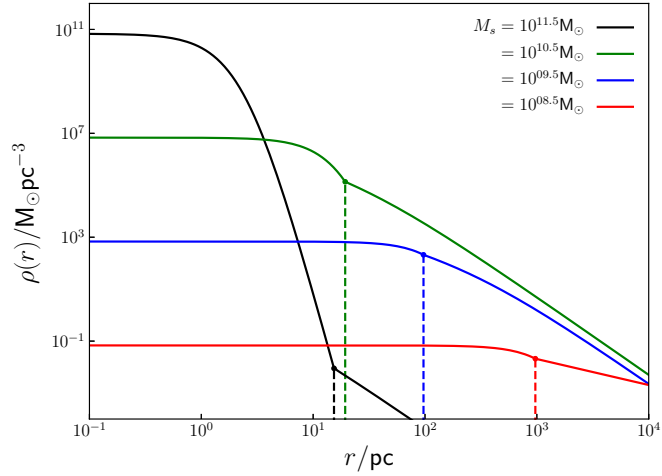}
     \caption{Examples of the density profiles from some of the selected
       configurations for Galaxy J0008-0004 obtained from the
       constraints in Fig.~\ref{fig:0004}. The core
       region is indicated by the plateau in the curves. Note that the transition to a NFW-like
       profile happens at larger radii for smaller central density. Shown are cases with the value $\alpha_{NFW} = 1$, except in the case $M_s =
       10^{8.5} \, M_{\odot}$ (red line)  for which
       $\log(\alpha_{NFW}) = -7$.
 	 } \label{fig:denplo}
\end{figure}

 {For brevity only representative results are shown, as in Fig.~\ref{fig:0004}, for the individual
cases of galaxies J0935-0003, J0008-0004 and J220329+020518; these cases include the
contribution of the luminous matter to the total mass of the
lens as in Eq.~\eqref{eq:24mod}.} For the purposes of clarity, in each figure we indicate the radius $r_s$ and total mass $M_s$ of the soliton
profile. We note that the free parameters $r_{\epsilon \ast}$ and $\alpha_{\rm
  NFW}$ appear well constrained if the soliton mass cannot provide the
total mass required by the lens system; in the examples shown, this
happens if $M_s < 10^{11.5} \, M_\odot$. Secondly, the credible regions for the 
parameters in
Fig.~\ref{fig:0004} are in agreement with the theoretical
expectations discussed in Sec.~\ref{sec:den}: there is a minimum
value for $r_{\epsilon \ast}$ due to the constraint imposed by
Eq.~\eqref{eq:16}, and a maximum value of $\alpha_{\rm NFW}$ appears
due to the maximal contribution of the NFW part of the profile to the
total mass in the lens, see also the
right panel in Fig.~\ref{fig:Lambdacr}. Likewise, notice that as
$\alpha_{\rm NFW} \to 0$ the value of the matching radius $r_{\epsilon
  \ast}$ is very well constrained, and this is easily understood from
Eq.~\eqref{eq:12}: it is $r_{\epsilon \ast}$ which determines alone the
contribution of the NFW part of the profile to the total
mass. Indeed, according to our parametrization in Sec.~\ref{sec:den} the NFW part of the density profile, under the limit $\alpha_{\rm NFW} \to 0$, becomes
\begin{equation}
\rho_{\rm NFW} (r) = \frac{\rho_s \, r_{\epsilon \ast}}{(r/r_s) \left(1+ r^2_{\epsilon \ast} \right)^8} \, .
\end{equation}
Apart from the presence of $r_{\epsilon \ast}$ (which in this case is bounded from above $r_{\epsilon \ast} \leq 1/\sqrt{15}$, see Eq.~\eqref{eq:8aa}), we also see that the behavior $1/r$ is the only one that survives from the NFW functional form, and then our results indicate that the outermost behavior $1/r^3$ is left unconstrained.

Finally, observe that the value $\log_{10}(M_s/M_\odot) = 7.5$
is excluded because the soliton mass $M_s$ is so small that the NFW part cannot compensate the required mass for the lens.  Recall that there is a matching (continuity) condition for the density profile in which the NFW density $\rho_{NFW}$ is always smaller than $\rho_s$, and this condition makes the NFW part of the profile unable to account for the total mass of the lens even in the limit $\alpha_{\rm NFW} \to 0$.

%{\bf remove this: this the code is not able to find any suitable values of the variables that could fit the data}. That is,

In summary, if the soliton is allowed to provide enough mass to
fulfill the matter contribution in the lens, say $M_s \sim 10^{11.5}
\, M_\odot$, the analysis will select large values for $r_{\epsilon
  \ast}$ so that the NFW tail contribution to the total matter is
minimal, see Eq.~\eqref{eq:12}. In contrast, if the soliton mass is
not large enough, $M_s <  10^{11.5} \, M_\odot$, it is then possible
to find appropriate pairs $(\alpha_{\rm NFW},r_{\epsilon \ast})$
for the NFW part of the profile to provide the needed mass for the
lens. In this respect, the striped credible regions in
Fig.~\ref{fig:0004} represent the degeneracy regions in the plane
$(\alpha_{\rm NFW},r_{\epsilon \ast})$ for the same mass
contribution of the NFW tail to the lens system.  {Thus, we can see
      that distinct credible regions can be found for the NFW
      parameters if the soliton mass is $10^{8.5} < M_s/M_\odot <
      10^{11.5}$, and that the constraints are in agreement with the
      semi-analytic analysis in Sec.~\ref{lensingQM}}.

Another quantity of interest is the resultant density profile of DM in
the lens system. Fig.~\ref{fig:denplo} shows examples of the density
profiles inferred from the posteriors of galaxy J0008-0004 in
Fig.~\ref{fig:0004} for a boson mass $m_{a22} = 1$. The soliton core
is clearly seen in all curves, and so too is the transition to the NFW part
of the profile.  {The corresponding matching radius
       $r_{\epsilon}$, in full units, is selected to be at $15.36$ pc,
       $19.34$ pc, $96.94$ pc and $969.4$ pc for the soliton masses
       $10^{11.5}M_{\odot}$, $10^{10.5}M_{\odot}$, $10^{9.5}M_{\odot}$
       and $10^{8.5}M_{\odot}$, respectively.} Not surprisingly, the largest core corresponds to the
configuration with the lowest soliton mass for which the matching
radius is close to the lower bound suggested in Eq.~\eqref{eq:8aa}.

\begin{figure*} 
  \centering
     \includegraphics[width=0.49\textwidth]{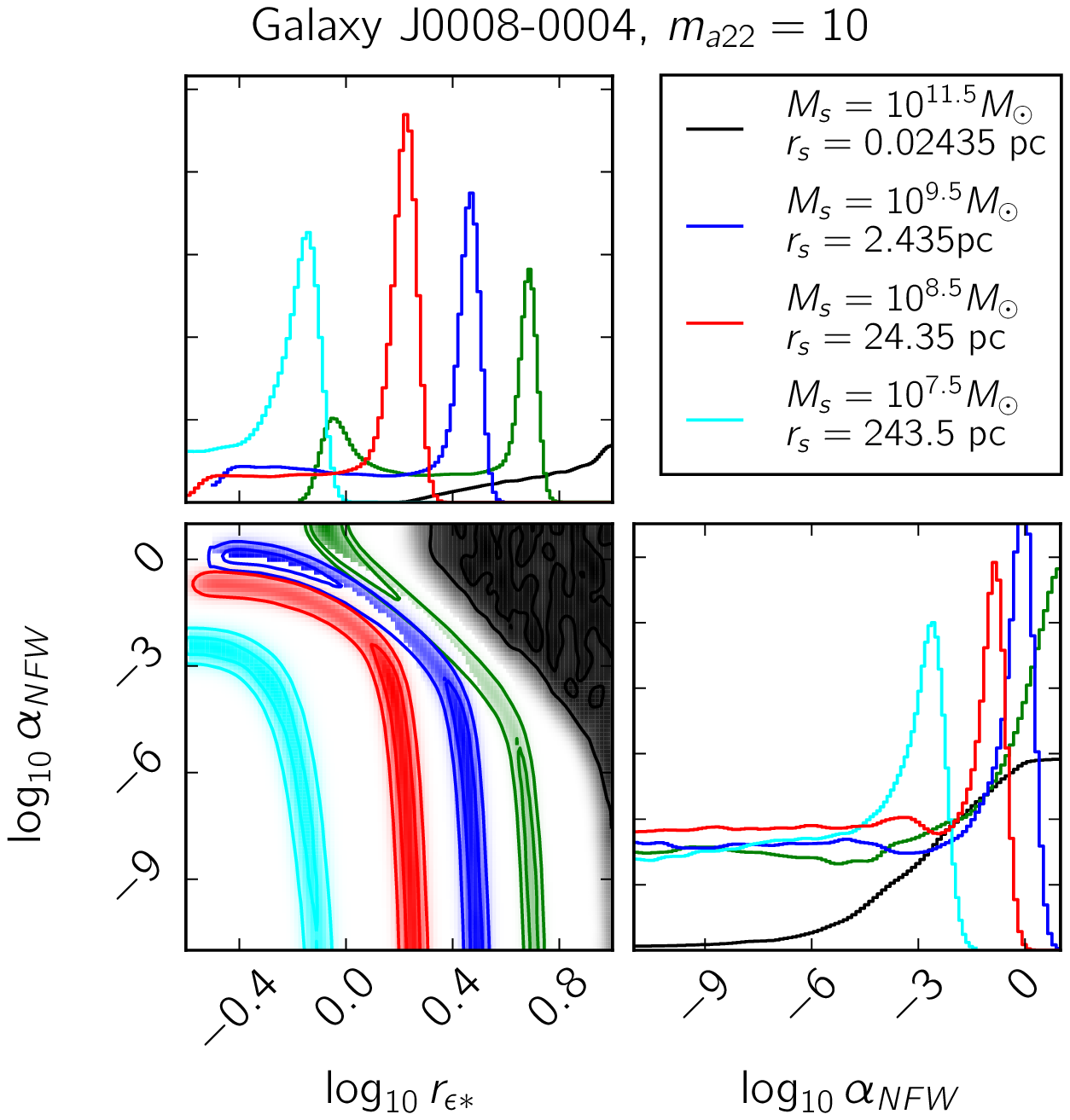}
    \includegraphics[width=0.49\textwidth]{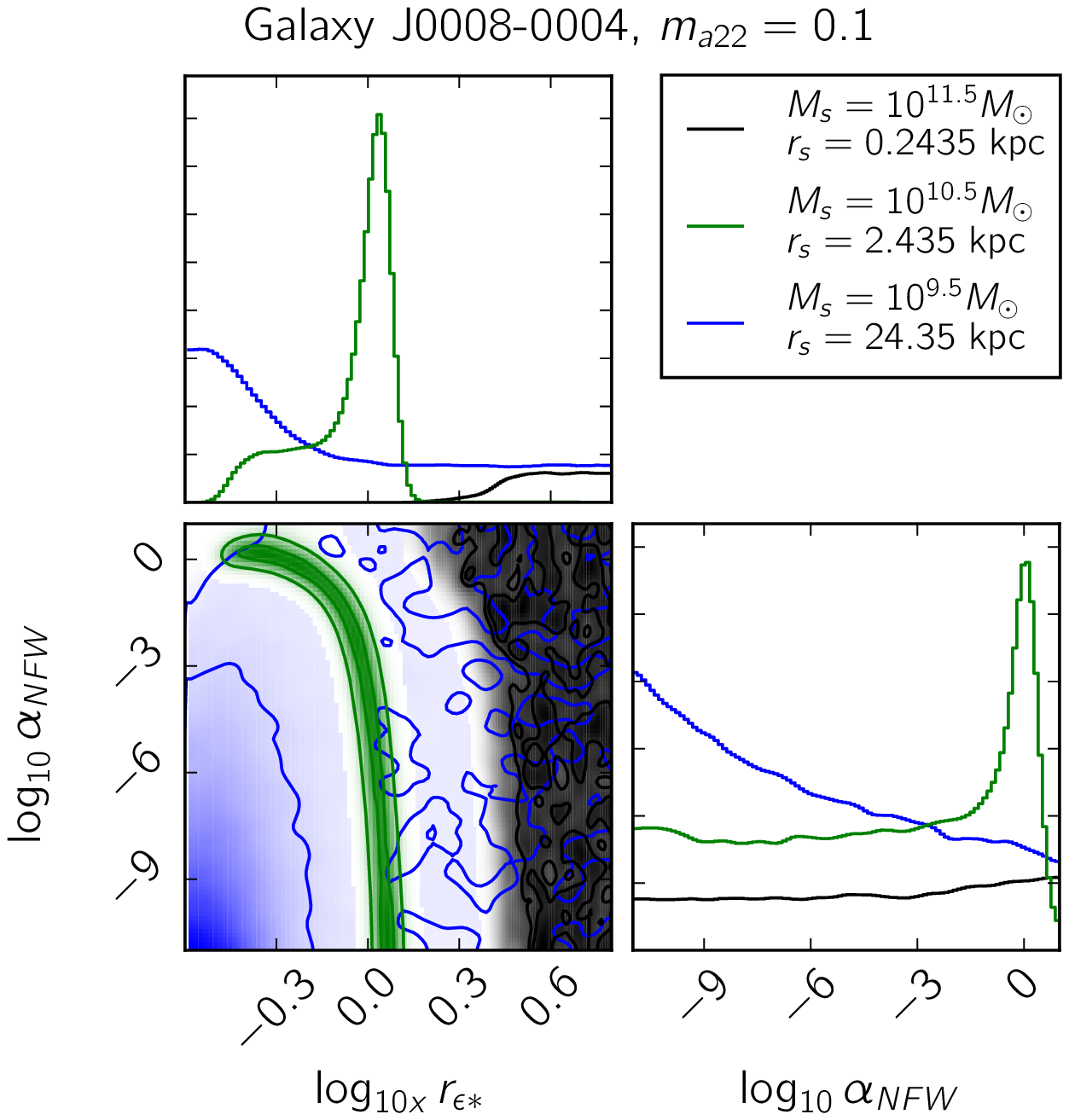}
    \includegraphics[width=0.49\textwidth]{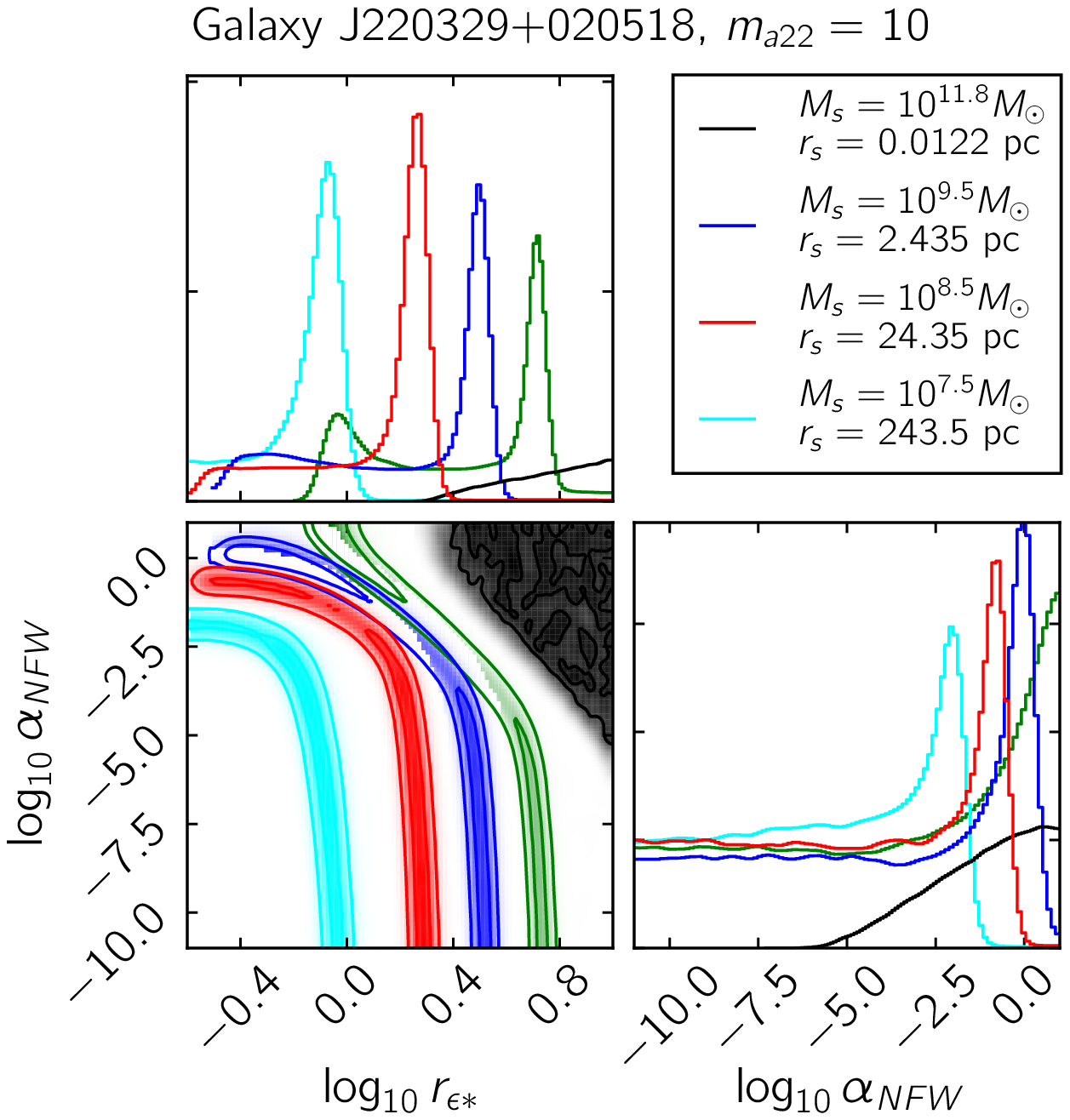}
    \includegraphics[width=0.49\textwidth]{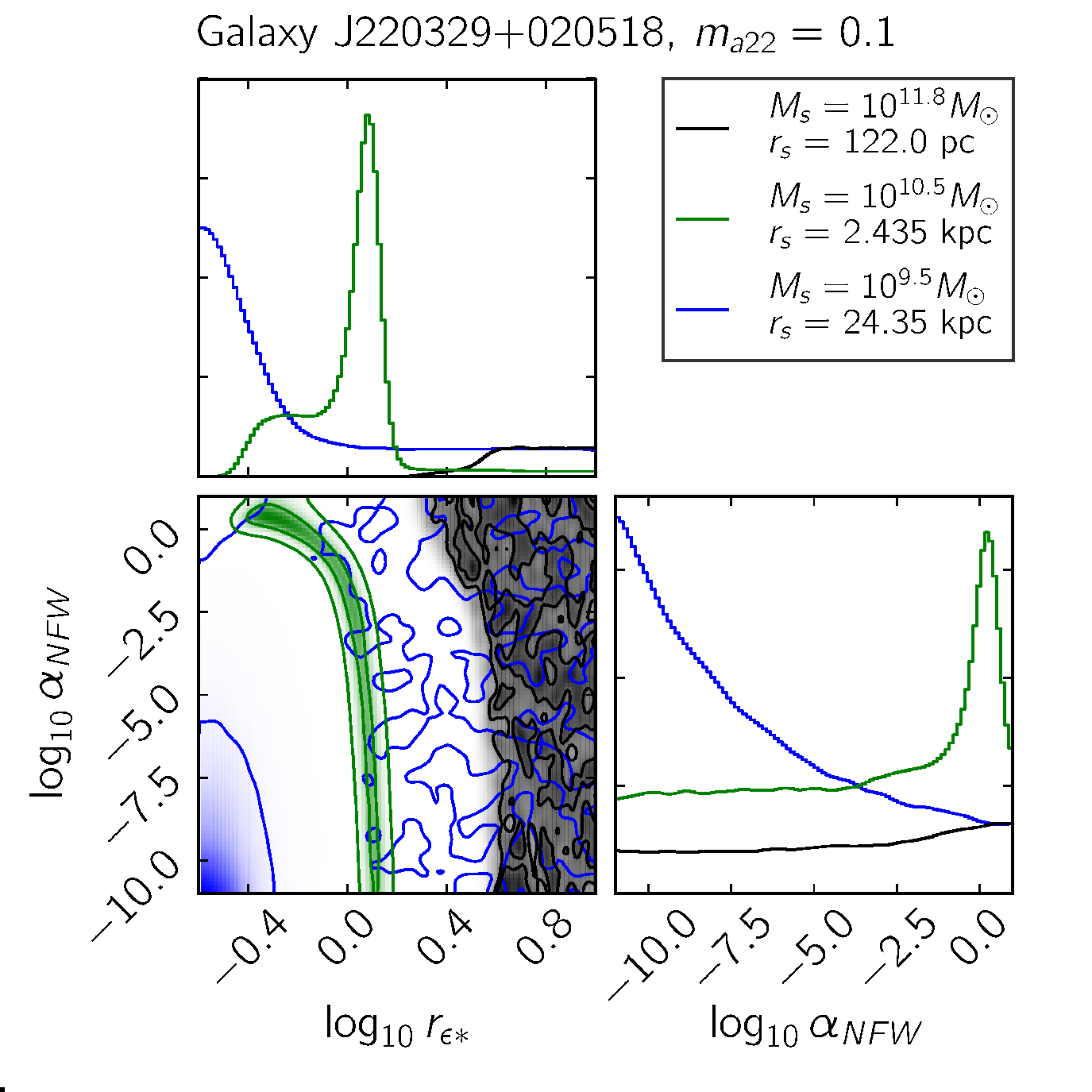}
    \caption{ {Same as figure \ref{fig:0004} but for galaxies J0008-0004 and J220329+020518. Here we show different boson masses: $m_{a22} =10$ (left panel) and $m_{a22} =0.1$ (right panel). See the text for more details.}}
%    \caption{The same type of triangle plot for the posteriors of galaxy J0008-0004 and J220329+020518, as in Fig.~\ref{fig:0004}, but nowconsidering different boson masses: $m_{a22} =10$ (left panel) and $m_{a22} =0.1$ (right panel). See the text for more details.}
 	\label{fig:masbos}
\end{figure*}

We also report in Fig.~\ref{fig:masbos} the results obtained for the
lens system J0008-0004  {and J220329+020518}, for larger or smaller values of the boson
mass.  {For the boson
      mass $m_{a22} = 10$ we obtain good constraints on the NFW
      parameters, but the soliton core is very compact in all cases, although a constraint cannot be found if $M_s = 10^{6.5} \, M_\odot$. For
      the boson mass of $m_{a22} = 0.1$, we can only obtain well
      defined constraints on the NFW parameters when the soliton mass
      is $M_s = 10^{10.5} \, M_\odot$, but not for larger or smaller
      values. Low values of $M_s$ imply values of the soliton
      radius $r_s$ that are larger than the Einstein radius, and this
      kind of cases are unable to satisfy the lensing constraints.} Hence, for a mass of $m_{a22} = 10$, the soliton is much more
compact, and it is not by itself adequate to describe a galaxy. But
given the fact that the parameters $\alpha_{\rm NFW}$ and $r_{\epsilon
  \ast}$ are also well constrained we conclude that the lensing effect
must be mostly attributed to the NFW part. This is not surprising, as
we had already indicated in Sec.~\ref{sec:grav} that strong lensing
could be achieved if $\alpha_{\rm NFW} \ll 1$. Moreover, a larger
boson mass is also in better agreement with recent cosmological
constraints~\cite[see][]{Irsic:2017yje} and with estimations based upon
satellite galaxies of the Milky Way and
Andromeda~\cite[see][]{Urena-Lopez:2017tob}.

In contrast, we can see that the constraints become more diffuse if we
consider a smaller boson mass of $m_{a22} = 0.1$, although there seems
to be some preference for the case in which $M_s = 10^{10.5}$, that
also corresponds to a larger soliton radius. This time the resultant
configuration would be in agreement with those found in the
statistical analysis carried out in~\cite{Gonzales-Morales:2016mkl}, which suggests that satellite galaxies put an upper bound on the boson mass that takes the form $m_{a22} < 0.4$.
\vspace{1cm}

\section{\label{sec:dis} Conclusions}
We have studied the properties of the so-called WaveDM density profile, assuming that it comprises the total DM contribution in galaxies for which a gravitational lens has been detected and measured. In doing so we have adapted the standard lens equations to the particular features of the WaveDM, in that we took into account its soliton core together with its NFW envelope, which is the complete form suggested by numerical simulations of cosmological structure under the WaveDM
hypothesis.

We then used the lens equations to make a comparison with actual
observations of some lens systems that seem to be DM dominated,
although we took into account their baryonic components in a
simplified manner. In carrying out the statistical analysis we considered carefully the role of the different free parameters of the WaveDM profile, and in particular the boson mass $m_a$ which has to be regarded as a fundamental parameter that should not vary from one galaxy to another.

The overall procedure was then to fix the value of the boson mass and the total mass within the soliton core in the configuration. In consequence, the soliton radius was fixed and the only free parameters were those of the NFW part of the density profile. In general terms, for large or small values of the boson mass, our results indicate that the soliton structure, if it is as massive as $10^{11.5} \, M_\odot$, is able to fit the measured Einstein radius in the lens systems studied, although this also requires the soliton structure to be
extremely compact when compared to the measured scales of the lensing galaxies. This result then indicates that galaxies in general cannot be explained by the soliton structure alone.

Because of the above, we had to consider the complete WaveDM density profile and constrain the NFW free parameters. Generically, and so far
for the cases we explored, our analyses suggest that the matching radius for the soliton and NFW parts of the profile is of the same order of magnitude as the soliton radius, $r_\epsilon \sim r_s$, which is in
agreement with the expectation from numerical simulations~\cite[e.g.][]{Schwabe:2016rze,Veltmaat:2016rxo,Mocz:2017wlg}. In
addition, the second free parameter is in general bounded from above as $\alpha_{\rm NFW} < 1$, which just means that the characteristic
NFW radius is larger than the soliton radius, $r_{\rm NFW} >
r_s$. Moreover, our results also suggest that the case $\alpha_{\rm NFW} \to 0$ is also possible, which in turn means that the density profile decays as $\rho \sim  r^{-1}$ at large radii.

On the other hand, for any given value of the boson mass, it was not
possible to constrain the NFW parameters in the case where the soliton
radius was larger than the Einstein radius, as in such cases the
soliton mass is insufficient to produce the required lensing
signal. Together with the aforementioned difficulty that the soliton
should not provide the whole mass of the lens, we can summarize our
results as $M_s/M_\odot < 10^{11.5}$ and $r_s < 6 \, \textrm{kpc}$. Given the similar masses and values of the Einstein radii in the selected sample of galaxies, these constraints can be taken as characteristic of the WaveDM model if the latter is considered to be the dark matter in them.

By means of Eq.~\eqref{eq:5}, the above inequalities can be combined in the following lower bound on the boson mass $m_a > 10^{-24}
\textrm{eV}$. Notice that this lower bound is in agreement with
previous constraints from cosmological and galactic scales,~\cite[see for
instance][]{Hlozek:2014lca,Chen:2016unw,Gonzales-Morales:2016mkl,Urena-Lopez:2017tob}. Although
the lens systems we considered are not able to put strong bounds on the
boson mass, they certainly indicate that most likely a complete WaveDM profile
(i.e. comprising a soliton core + NFW tail) is necessary to account for 
all the diverse observations at galaxy scales.

As a final note, the lens systems studied here have a subdominant,
although non-negligible, baryonic contribution. We expect to extend
our analysis to a larger sample considering other surveys with a more detailed and specific description of the baryonic matter contained that could give us better constraints on the soliton
features. This is ongoing work that will be presented elsewhere.
\acknowledgments
LAU-L wishes to thank Andrew Liddle and the Royal
Observatory, Edinburgh, for their kind hospitality in a fruitful
sabbatical stay. This work was partially supported by Programa para el
Desarrollo Profesional Docente; Direcci\'on de Apoyo a la
Investigaci\'on y al Posgrado, Universidad de Guanajuato, research Grant 206/2018; Programa Integral de Fortalecimiento Institucional;
CONACyT M\'exico under Grants No.~232893 (sabbatical), No. 167335,
No. 179881, No. 269652, No.182445 and Fronteras 281; Fundaci\'on Marcos Moshinsky; and the Instituto Avanzado de Cosmolog\'ia Collaboration.

\appendix
\section{Integral solutions \label{sec:integral-solutions-}}
Some useful analytical solutions are given here for the integrals in
Eq.~\eqref{eq:8a}. For the first branch $\theta_\ast < r_{\epsilon
  \ast}$ the formula for the first integral is
  \begin{equation}
    \label{eq:15}
    \int \limits^{\sqrt{r^2_{\epsilon \ast}-\xi^2_\ast}}_0 \frac{dz}{\left[ 1
        + \alpha^2_{\rm sol} \, \hat{r}^2 \right]^8} =
    \frac{\alpha^{-1}_{\rm sol}}{(1 + \alpha^2_{\rm sol}
      \xi^2_\ast)^{15/2}} \int \limits^x_0 \cos^{14} u \, du \, , \quad
    \tan x = \alpha_{\rm sol} \left( \frac{r^2_{\epsilon \ast} -
        \xi^2_\ast}{1 + \alpha^2_{\rm sol} \xi^2_\ast} \right)^{1/2} \,
    ,
  \end{equation}
  where
  \begin{eqnarray}
    \label{eq:13}
    \int \limits^x_0 \cos^{14} u \, du = \frac{429}{2048} x +
    \frac{1001}{16384} \left[ 3 \sin(2x) + \sin(4x) + \frac{1}{3}
      \sin(6x) + \frac{1}{11} \sin(8x) + \frac{1}{55} \sin(10x) \right. 
      \nonumber \\ 
      \left.+ \frac{1}{429} \sin(12x) + \frac{1}{7007} \sin(14x) \right] \, ,
  \end{eqnarray}
whereas for the second integral we obtain
  \begin{equation}
\int \limits^\infty_{\sqrt{r^2_{\epsilon \ast} - \xi^2_\ast}}
\frac{dz}{\hat{r} \left( 1+ \alpha_{\rm NFW} \, \hat{r} \right)^2} = 
\begin{dcases}
\frac{1}{x^2-1}\left( 1- \frac{\sqrt{y^2
	-x^2}}{1+y} - \frac{2 \arctanh}{\sqrt{1-x^2 }}{
	\left[\frac{\sqrt{1-x^2}}{1+y+\sqrt{y^2-x^2}}\right]} \right) & x < 1 \, , \\
\frac{1}{3}\left( 1- 
\frac{y+2}{y+1}\sqrt{\frac{y-1}{y+1}}
\right) & x = 1 \, , \\
\frac{1}{x^2-1}\left( 1- \frac{\sqrt{y^2
	-x^2}}{1+y} - \frac{2 \arctan}{\sqrt{x^2-1 }}{
	 \left[\frac{\sqrt{x^2-1}}{1+y+\sqrt{y^2-x^2}}\right]} \right) & x > 1 \, .
    \end{dcases} \, , \label{eq:9}
\end{equation}
where $ x = \alpha_{\rm NFW} \xi_\ast$ and $y = \alpha_{\rm
  NFW}r_{\epsilon \ast} $. By setting $y = x$, which is equivalent to
$r_{\epsilon \ast} = \xi_\ast$, in Eq.~\eqref{eq:9} we obtain the
solution for the second branch in Eq.~\eqref{eq:8a}. For the case
$\xi_\ast = 0$, which is used in Eq.~\eqref{eq:20}, the integral
result simply is
\begin{equation}
     \int \limits^\infty_{r_{\epsilon \ast}} \frac{dz}{z \left( 1 +
            \alpha_{\rm NFW} \, z \right)^2} = \ln \frac{\left( 1 + \alpha_{\rm NFW} \, r_{\epsilon \ast}
            \right)}{ \alpha_{\rm NFW}r_{\epsilon \ast}}- \frac{1}{\left( 1 + \alpha_{\rm NFW} \, r_{\epsilon \ast}
                        \right)} \, .
  \end{equation}

\bibliography{bibli}

\end{document}